\documentclass[fleqn,usenatbib]{mnras}

\usepackage{newtxtext,newtxmath}
\usepackage{siunitx}
\usepackage{lipsum}
\usepackage{diagbox}
\usepackage[T1]{fontenc}

\DeclareRobustCommand{\VAN}[3]{#2}
\let\VANthebibliography\thebibliography
\def\thebibliography{\DeclareRobustCommand{\VAN}[3]{##3}\VANthebibliography}

\usepackage{graphicx}	% Including figure files
\usepackage{amsmath}	% Advanced maths commands
\usepackage{tablefootnote}
\usepackage{subcaption}
\usepackage{xcolor}
\usepackage{hyperref}
\newcommand{\halpha}{\mbox{H$\alpha$}}
\newcommand{\hbeta}{\mbox{H$\beta$}}

\newcommand{\oII}{\mbox{O{\sc ii}}}
\newcommand{\oIII}{\mbox{O{\sc iii}}}

\newcommand{\nII}{\mbox{N{\sc ii}}}

\newcommand{\Msun}{\mbox{ M$_\odot$}}

\title[Getting to know the Stellar Clusters in NGC~1569]{Getting to know the Stellar Clusters in NGC~1569: Bayesian inference of stellar cluster properties in a dwarf starburst galaxy}

\author[Bjarki Björgvinsson]{
Bjarki Björgvinsson,$^{1}$\thanks{E-mail: bjarki.bjorgvinsson@durham.ac.uk} Anna F. McLeod, $^{1, 2}$ Bronwyn Reichardt Chu,$^{1, 2}$ \newauthor Magdalena J. Hamel-Bravo,$^{3, 4}$ Deanne B. Fisher,$^{3, 4}$ Mark R. Krumholz,$^{4, 5}$
\\
$^{1}$Centre for Extragalactic Astronomy, Department of Physics, University of Durham, South Road, Durham DH1 3LE, UK\\
$^{2}$Institute for Computational Cosmology, Department of Physics, University of Durham, South Road, Durham DH1 3LE, UK\\
$^{3}$Centre for Astrophysics and Supercomputing, Swinburne University of Technology, Hawthorn, Victoria 3122, Australia\\
$^{4}$ARC Centre of Excellence for All Sky Astrophysics in 3 Dimensions (ASTRO 3D), Australia\\
$^{5}$Research School of Astronomy and Astrophysics, Australian National University, Cotter Rd., Weston ACT 2612, Australia
}

\date{Accepted XXX. Received YYY; in original form ZZZ}

\pubyear{2026}

\begin{document}
\label{firstpage}
\pagerange{\pageref{firstpage}--\pageref{lastpage}}
\maketitle

% Abstract of the paper
\begin{abstract}
We present a Bayesian analysis of star clusters in the dwarf starburst galaxy NGC 1569 based on high-resolution \textit{Hubble Space Telescope} imaging combined with integral-field spectroscopy from the \textit{Keck Cosmic Web Imager}, obtained as part of the DUVET survey. For each cluster identified, we infer posterior probability distributions for mass and age using a forward modelling method that properly accounts for uncertainties due to stochastic sampling of the IMF. We investigate how the inferred properties depend on photometric coverage by repeating the analysis with different filter combinations, including mock extensions to the ultraviolet and near-infrared that emulate the addition of \textit{HST} UV bands and \textit{James Webb Space Telescope} imaging. We find that, while inclusion of these wavelength regimes further breaks age and mass degeneracies, the currently available data yields reasonably strong constraints on cluster parameters. We compare inferred cluster properties to the conditions of the local interstellar medium, and find evidence for multiple interesting correlations. The truncation mass of the cluster mass function varies with galactocentric distance, particularly moving off the disk, consistent with a dependence on the density of the interstellar medium. Cluster mass positively correlates with metallicity, suggesting that massive clusters preferentially form in pre-enriched gas, and the ionisation state of the gas, reflecting the increased prevalence of high-mass stars in high-mass clusters. These results demonstrate the power of Bayesian, initial mass function-aware modelling for resolving cluster populations in nearby starburst dwarfs and provide new insight into how cluster formation and feedback respond to local galactic conditions.
\end{abstract}
%The abstract should briefly describe the aims, methods, and main results of the paper.
%It should be a single paragraph not more than 250 words (200 words for Letters).
%No references should appear in the abstract.

% Select between one and six entries from the list of approved keywords.
% Don't make up new ones.

%%%%%%%%%%%%%%%%%%%%%%%%%%%%%%%%%%%%%%%%%%%%%%%%%%

%%%%%%%%%%%%%%%%% BODY OF PAPER %%%%%%%%%%%%%%%%%%

\section{Introduction}\label{Introduction}
Dwarf galaxies are the most common kind of galaxies in the Universe \citep{2006A&A...459..745F, 2015A&A...575A..96G}, and due to their shallow potential wells they are highly sensitive to feedback processes \citep{Emerick_2018, Collins_2022, Marasco_2023}. This means that stellar feedback from massive stars (M$\gtrsim8\Msun$), including winds, ionising radiation, and supernovae (SNe) can more easily drive outflows and shape the structure of the interstellar medium (ISM) in lower mass galaxies, regulating future star formation. Due to the combination of sensitivity to feedback and low gas masses, dwarf galaxies can blow out much of their available gas leading to bursty star formation histories (SFHs) characterised by periods of high specific star formation rate (sSFR) and strong galactic winds followed by more quiescent periods, when the gas available for star formation is mostly depleted \citep{Emami_2019, furlanetto_bursty_2022}. Star formation in these galaxies is often compared to that in high-redshift galaxies due to the similarties in physical conditions, such as low metallicity, bursty star formation, and irregular morphology \citep{ostlin_ly_2014, 2021A&A...646A.138I, flores_local_2021}.

Because a large fraction of stars form in clustered environments \citep{lada_embedded_2003,portegies_zwart_young_2010, krumholz_star_2019, girichidis_physical_2020}, studies targetting star clusters allow for highly detailed studies of the effects of feedback, making them an excellent laboratory for linking star formation and feedback effects. 
%This is particularly true in dwarf starbursts, where the intense star formation together with the shallower potential well make stellar feedback effects more clear as the ISM is driven by young, massive stars.
The cluster mass function (CMF) provides an example of how one can make such links. The CMF of young star clusters is typically described by a powerlaw, but it may exhibit truncation at a critical mass $M_c$, which encodes information about the physical conditions of the ISM at the time of cluster formation \citep{2009A&A...494..539L, 2018MNRAS.473..996M, 2020SSRv..216...69A, 2022ApJ...928...15W, della_bruna_stellar_2022, tang_cluster_2024}. In particular, theoretical models predict that $M_c$ depends on environmental properties such as gas surface density, pressure, and turbulence, as well as on the efficiency and timing of stellar feedback \citep{reina-campos_unified_2017}. By studying the properties of star clusters in dwarf starbursts (e.g. age, mass, spatial distribution), where star formation is highly clustered and feedback effects are expected to be particularly strong, spatially resolved cluster mass measurements enable tests of whether $M_c$ is regulated by gas density, turbulence, or feedback-driven disruption. Similarly, spatially mapping the age of the clusters allows us to examine how star formation progressed across the galaxy, i.e. are the starbursts localised, or global across the entire galaxy.

Tests of this sort require that we be able to resolve individual clusters in distant galaxies. Using high-resolution HST photometry such resolution up to a distance of $\sim10$ Mpc. We can then supplement such data by the addition of integral field unit (IFU) spectroscopy, which offers stronger constraints by letting us define custom photometric filters that capture specific spectral features useful in discriminating cluster properties. IFU data also allow us to probe local nebular conditions, which can be used to quantify the effects of feedback from stellar clusters.

However, studies of this type face a fundamental challenge: when studying star clusters, it is common to use simple stellar populations (SSPs) to constrain the properties (e.g. mass and age) of unresolved clusters \citep{1999ApJS..123....3L, 2005ApJ...621..695V, 2009ApJ...699..486C, 2010ApJ...708...58C}. Assuming a cluster age, initial mass function (IMF), and stellar evolutionary tracks, the luminosity of the synthetic cluster can be obtained and compared with that of the observed clusters. However, this starts to fail at masses lower than $\sim10^4$\Msun\ \citep{krumholz_star_2014, krumholz_slug_2015, stanway_exploring_2023}, where clusters are unlikely to fully sample the initial mass function (IMF), and the relationships between photometry and physical properties become non-deterministic. Thus, in order to conduct accurate studies of star clusters, it is necessary to account for the stochastic nature of star formation.

Many stellar population synthesis codes have been developed to tackle this problem. Here we use the Stochastically Lighting Up Galaxies code, \verb|SLUG| \citep{silva_slugstochastically_2012, 2014MNRAS.444.3275D, krumholz_slug_2015}, to infer the properties of the clusters in the nearby (3.25 Mpc, \citealt{tully2013}), low metallicity (12+log(O/H) = 8.22, \citealt{MagdaPaper}), low mass (log(M$_\star$/\Msun) = 8.6, \citealt{Leroy2019}) galaxy NGC~1569. This galaxy is undergoing an intense starburst, possibly triggered by gravitational interactions with a passing HI cloud or members of its galaxy group \citep[IC 342;][]{1998A&A...337...64S, 2002A&A...392..473S, 2012AJ....144..134H, 2013AJ....145..146J}. It contains two supermassive star clusters (SSCs), SSC-A and SSC-B \citep{hunter_star_2000}, with \halpha-bright bubbles and filaments \citep{westmoquette_gemini_2007}, and a high-metallicity region surrounding SSC-B \citep{MagdaPaper}. These features point to strong stellar feedback from the SSCs.
The star formation history of NGC~1569 has been studied in the past, using many different methods (see for example \citealt{Greggio_1998, anders_star_2004, angeretti_complex_2005}), but these works have all assumed that each cluster fully samples the IMF which may lead to inaccurate age measurements \citep{krumholz_slug_2015}. Our goal here is therefore to revisit the cluster properties in NGC 1569 using modern Bayesian inference methods that properly account for these effects, and then to use the resulting inferred cluster properties to characterise the nature of star formation in this starburst.

This paper is structured as follows: in Section~\ref{sec:Data} we describe the extraction of cluster photometry from observational data as well as the parameters used to generate the synthetic star cluster library with \verb|SLUG|. In Section \ref{sec:FilterSelection} we outline our method to infer cluster properties from photometry, and describe how we choose which photometric filters to use in this analysis. We present our results in Section~\ref{sec:Analysis}, plotting cluster masses and ages against each other as well as investigating whether there are spatial correlations or correlations between cluster and ISM properties. In Section~\ref{sec:Discussion} we discuss these results and their implications for cluster formation, and in Section~\ref{sec:Summary} we summarise the key results.

\section{Data}\label{sec:Data}
\subsection{Observations}\label{sec:Obs}
\subsubsection{KCWI integral field spectroscopy}
\begin{figure*}
    \centering
    \includegraphics[width=1\linewidth]{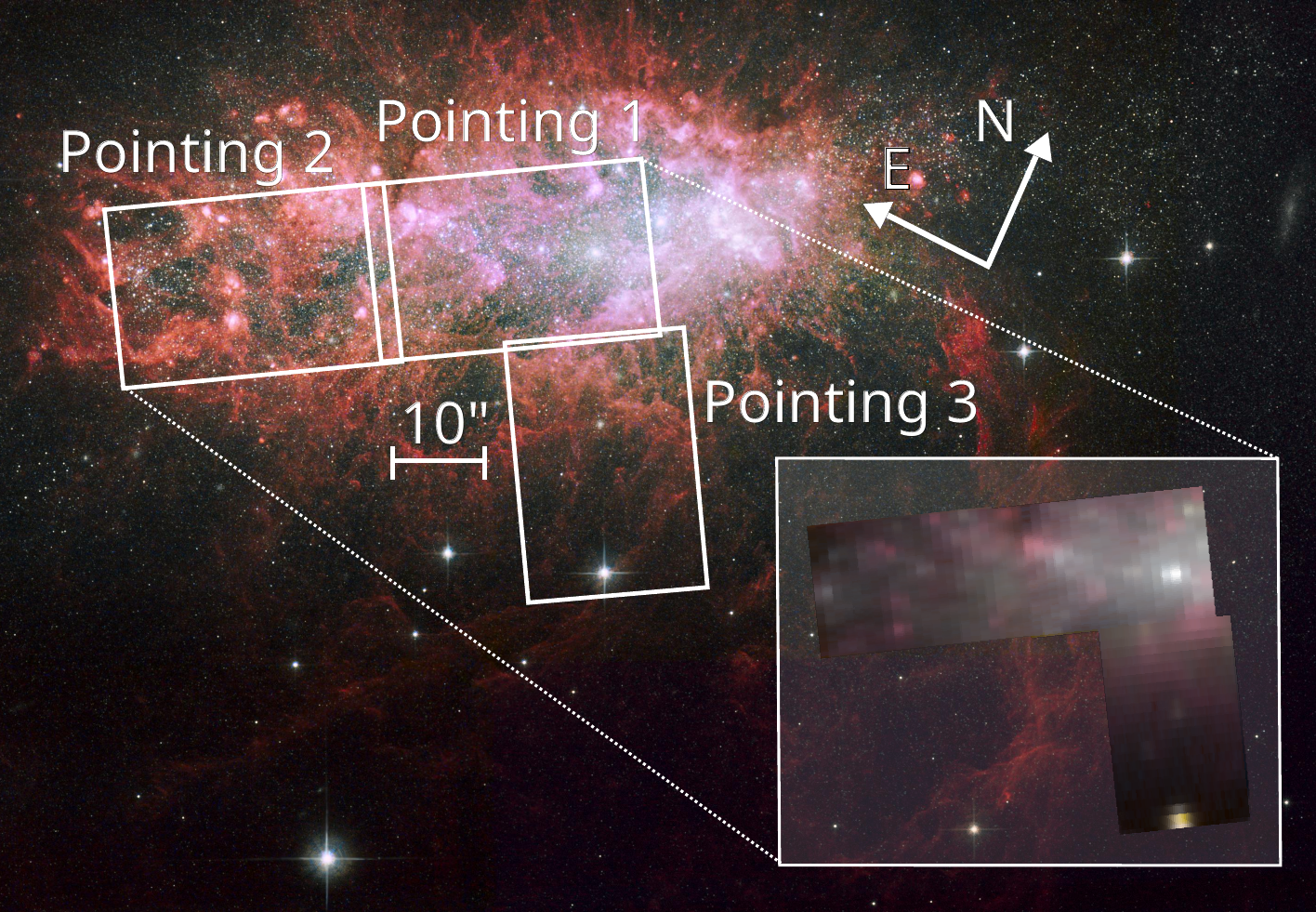}
    \caption{HST composite image of NGC~1569; red: F658N filter (\protect\halpha\ + [\protect\nII]), green: F606W filter, light blue: F502N filter ([\protect\oIII]) and dark blue: F487N filter (\protect\hbeta). The three rectangles show our three KCWI positions. The KCWI data cover $\protect\SI{54}{\arcsecond}$($\protect\sim$800 pc) across the major axis and $\protect\SI{48}{\arcsecond}$($\protect\sim$700 pc) across the minor axis. The inlay in the bottom right shows a 3 colour image constructed from our KCWI data using our three broadbands as RGB, (red: KCWI 4601, green: KCWI 4101, blue: KCWI 3600). See also Fig.~1 in \protect\cite{MagdaPaper}}
    \label{fig:NGC1569}
\end{figure*}

NGC~1569 was observed using three pointings of Keck/KCWI \citep{morrissey_keck_2018}, as part of the Deep near-UV observations of Entrained gas in Turbulent galaxies (DUVET) survey \citep{cameron_duvet_2021}. Two pointings observed the disk east of SSC A, but the third was aimed at an \halpha\ filament aligned with the minor axis. The coverage of the pointings is shown in Fig.~\ref{fig:NGC1569}. The observational strategy and data reduction of DUVET targets are described in \cite{cameron_duvet_2021} and \cite{mcpherson_duvet_2023}, while details on the observations and data reduction of NGC~1569 are available in \cite{MagdaPaper}, and are summarised below.

Observations were conducted on November 14, 2020, with a mean seeing of \SI{1.2}{\arcsecond}. Using two different configurations of the blue medium-dispersion grating, the ``blue'' setting, centered at 4050~\AA\ (3600 - 4498~\AA) and the ``red'' at 4700~\AA\ (4256 - 5135~\AA). This gives a continuous wavelength range of 3600 - 5130~\AA, with a spectral resolution of R$\sim2000$. Since our analyses are based on  photometry (see Section~\ref{sec:Analysis}), we split the cube into three equal bandpasses of 500~\AA\ each. This gives us custom bands covering 3600 - 4100~\AA, 4101 - 4600~\AA\ and, 4601 - 5100~\AA. In the rest of this work we refer to these bands by their starting wavelengths. In addition, we define two narrowbands centred on the emission lines [\oIII]$\lambda$5007 and \hbeta, each 5~\AA\ wide, as well as one 20~\AA\ band containing no strong emission lines at 4925~\AA\ - 4945~\AA, to allow for continuum normalisation of the [\oIII] and \hbeta\ lines.

All observations used the large IFU slicer, resulting in a field of view of $\SI{20}{\arcsecond}\times\SI{30}{\arcsecond}$ and a spaxel size $\SI{0.29}{\arcsecond}\times\SI{1.35}{\arcsecond}$ per pointing. After resampling the spaxels with \verb|Montage|\footnote{\url{http://montage.ipac.caltech.edu/}}, the spaxels have size $\SI{0.87}{\arcsecond}\times\SI{0.87}{\arcsecond}$, approximately half the FWHM of the PSF.

At 3.25~Mpc, the distance to NGC~1569, $\SI{0.87}{\arcsecond}$ corresponds to a physical size of $\sim$13.7 pc. This means that our spatial sampling is much larger than the expected size of star clusters, which have a mean size of 2-3 pc in most galaxies \citep{ryon_effective_2017, brown_radii_2021}. Additionally, the separation between clusters is often smaller than the FWHM of the PSF, causing blending of multiple clusters in the KCWI images. Given these limitations, we cannot identify individual clusters in the KCWI data. To remedy this, we use co-spatial high-resolution photometry from HST.

\subsubsection{HST imaging}
NGC~1569 was observed using the ACS/WFC broad-band filters F606W (4634 - 7180~\AA) and F814W (6870 - 9632~\AA) by \citet{2006hst..prop10885A}, proposal ID 10885. These data have a pixel size of $\SI{0.05}{\arcsecond}$ and a PSF of $\lesssim \SI{0.1}{\arcsecond}$, which corresponds to spatial scales of $\sim 0.79$ pc and $\sim 1.6$ pc respectively, so most clusters are easily resolved and larger than one pixel. Including the HST photometry in the analysis also serves to increase the wavelength coverage of the KCWI data to redder wavelengths, extending it to 9632~\AA. This allows us to obtain more information on low-mass stars as well as younger, more embedded clusters, which are difficult to detect in blue wavelengths, particularly in the absence of massive stars, and in regions of high extinction.

\subsection{Photometry}
\subsubsection{Observed photometry}\label{sec:observed-photometry}
As mentioned above, the high spatial resolution HST images are used to deblend clusters in the KCWI data by using HST data to centroid the cluster coordinates,  then extracting photometry for those coordinates from the KCWI data. Our approach is analogous to those described by \cite{PampelMuse} and \cite{mcleod_stellar_2020}, who use high-resolution photometry to resolve individual massive stars within stellar clusters in order to extract their spectra in lower resolution IFU data. 

\begin{figure*}
    \centering
    \includegraphics[width=\linewidth]{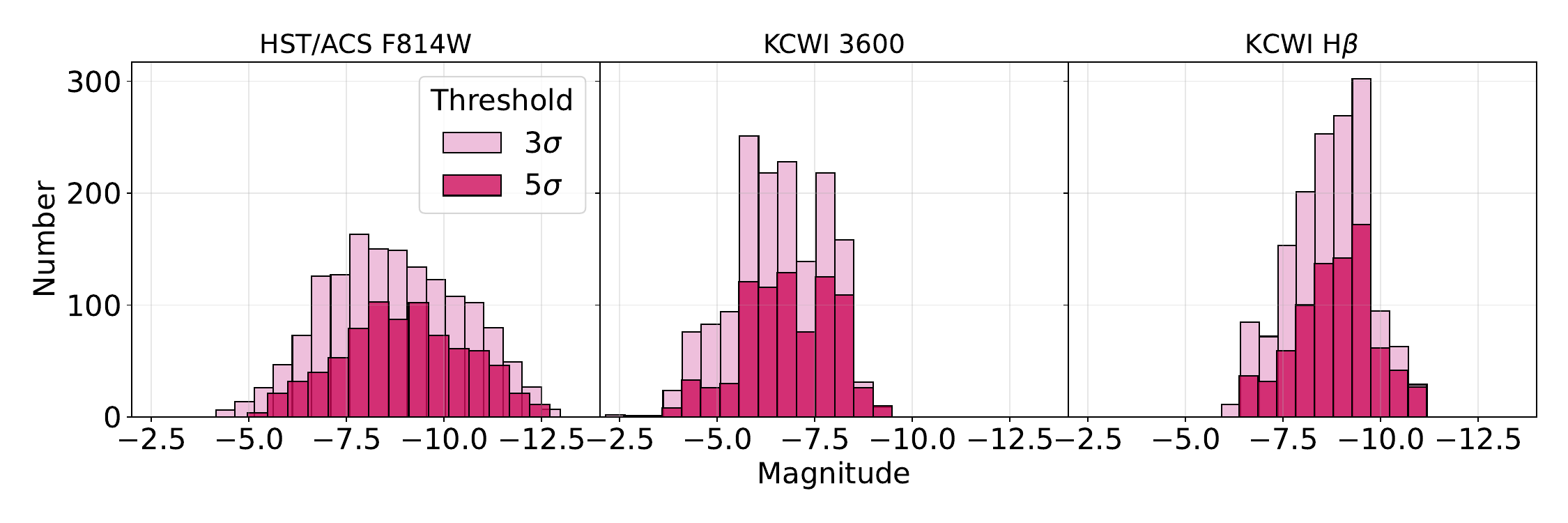}
    \caption{Histograms showing the distribution of magnitudes in 3 bands: HST ACS F814W, KCWI 3600, and KCWI \protect\hbeta\ using both the 3$\sigma$ (pink) and 5$\sigma$ (magenta) detection thresholds. The magnitudes have been corrected for foreground extinction.}
    \label{fig:mag-hist}
\end{figure*}

Our method is as follows. First, we perform aperture photometry to obtain the flux in both HST filters using \verb|photutils| \citep{larry_bradley_2024_12585239}, which allows for background subtraction and convolution of an image with a Gaussian kernel before the detection takes place. The parameters for background subtraction, Gaussian convolution, and subsequent source detection are shown in Table \ref{tab:photutils}.
Changing the detection threshold significantly affects how many sources are identified. Using a threshold of $3\sigma$ results in 2300 identified sources in F606W and 2074 in F814W, while using $5\sigma$ results in 1147 and 1159 sources, respectively. We adopt the $3\sigma$ catalogue for the remainder of this work in order to maximise completeness, while subsequent photometric quality cuts remove spurious detections.
We then cross-match the two source lists using \verb|TOPCAT| \citep{topcat}, to make a list that contains only clusters detected in both filters, where the distance between the detected centroids is limited to $<1$ arcsec. This corresponds to $\sim 15$ pc at the distance of NGC~1569, large enough to account for different morphologies of clusters in different wavelengths while being small enough to not cause confusion between sources that are only detected in one of the two filters. This leaves us with 1688 sources that have detections in both filters. For completeness, this corresponds to 859 sources if using the 5$\sigma$ detection threshold. A comparison of magnitude distributions in F814W, using the two detection thresholds is shown in the left-most panel of Fig.~\ref{fig:mag-hist}. 

\begin{table*}
    \caption{Description and motivation of the parameters used for point source detection and aperture photometry using \texttt{photutils}.}
    \begin{tabular}{|l|l|l|}
    \hline
         \textbf{Parameter} & \textbf{Value} & \textbf{Description}\\
         \hline
         box\_size & (50,50) & \begin{tabular}{@{}l@{}}The size of the box used to estimate the local background at each pixel, in units of pixels. Chosen to be\\ significantly larger than the largest source, while still estimating the bakground locally. \end{tabular}\\
         \hline
         filter\_size & (3,3) & \begin{tabular}{@{}l@{}}The size of the median filter applied to the background map. The value is set to ensure our background\\ is not overestimated due to extended sources without smoothing out all background variation.\end{tabular}\\
         \hline
         bkg\_estimator & SExtractorBackground & \begin{tabular}{@{}l@{}}The method used to estimate the background value. The SExtractorBackground estimator uses the formula\\ for background estimation used in \verb|SExtractor| \citep{SExtractor}$^*$.\end{tabular}\\
         \hline
         fwhm & 3 & \begin{tabular}{@{}l@{}}The FWHM of the Gaussian kernel used in the convolution of the image. Set to roughly match the PSF of \\the HST images to smooth out any bright point sources.\end{tabular}\\
         \hline
         size & 5 & \begin{tabular}{@{}l@{}}The size of the Gaussian kernel used in the convolution of the image. Chosen to be slightly larger than the \\FWHM as to not blend out smaller sources, while also preventing sources from blending together.\end{tabular}\\
         \hline
         npixels & 4 & \begin{tabular}{@{}l@{}}The minimum number of pixels required for a source to be detected in the convolved image. 4 pixels \\ corresponds roughly to a round cluster with a diameter of $\sim 1.4$ pc. After convolving, smaller sources may \\ appear a few pixels larger, but the total flux remains the same.\end{tabular}\\
          \hline
         thresh & 3 & \begin{tabular}{@{}l@{}}The threshold flux a pixel must have to be counted as part of a source. The value corresponds to a \\factor of the local background RMS, i.e. the flux of the source must be 5 $\sigma$ higher than the background.\end{tabular}\\
         \hline
    \end{tabular}
    \hbox{\hspace{-2.8cm}$^*$Background is equal to (2.5$\times$ median) - (1.5$\times$ mean) in the box. Unless (mean - median)/$\sigma_{bg}$ > 0.3, in which case the median is used}
    \label{tab:photutils}
\end{table*}

\begin{figure}
    \centering
    \includegraphics[width=1\linewidth]{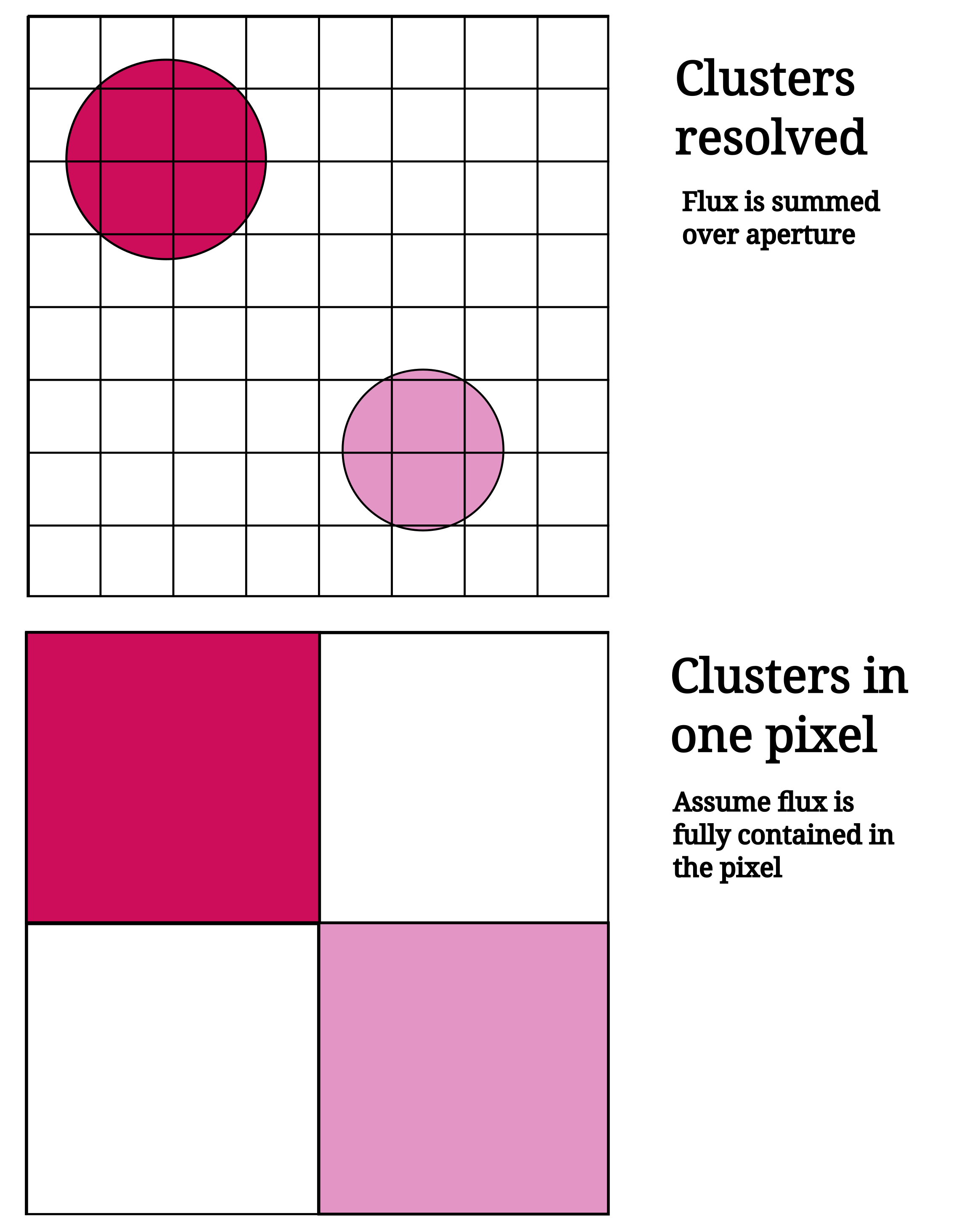}
    \caption{A schematic comparing the methods of aperture photometry extraction in the the high spatial resolution HST (upper panel) and lower resolution custom KCWI (lower panel) filters.}
    \label{fig:schematic}
\end{figure}

Second, using the HST source coordinates, we then extract the photometry from the KCWI data. We begin this process by checking WCS alignment of the KCWI images with respect to HST by convolving the HST F606W image to the resolution of the KCWI data and performing a visual inspection by blinking the images to ensure that corresponding sources in the images coincide. 
Due to crowding and the native size of the KCWI pixels, the clusters are poorly resolved and highly blended. We attempt to fit the PSF, but find that the field is too crowded for this to be effective. To obtain KCWI fluxes, we therefore resort to the simple assumption that all the flux from a single cluster is localised in a single pixel of the KCWI image, as illustrated in Fig.~\ref{fig:schematic}. Under this assumption, the flux that we attribute attribute to each cluster in each of the KCWI custom filters is simply the flux in the pixel that contains the cluster centroid. This approach implicitly assumes that each pixel contains negligible flux from neighbouring pixels, and that each pixel is dominated by the light of the cluster within it. This approach also necessitates that we drop all pixels that contain more than one cluster centroid, as clearly we cannot separate the light of the two clusters. This cut removes only 8 sources, but even in cases where centroids do not fall into the same pixel, there may be bleeding and blending from neighbouring sources. We also remove from further analysis any clusters that fall on the edges of our KCWI data, where blending is worse; this further reduces our number of sources to 1115. 

Our third and final step, now that we have extract photometry for both HST and KCWI filters, is to correct for foreground Milky Way extinction. We do so using \verb|pyneb| \citep{pyneb}, with the \cite{1989ApJ...345..245C} extinction law and an extinction of $A_V = 1.85$ \citep{2011ApJ...737..103S}. The resulting extinction-corrected magnitudes will form the basis for the analysis we present in the remainder of this paper. The extinction-corrected magnitude distribution in KCWI 3600 and KCWI \hbeta\ for the remaining sources is shown in the middle and right panels of Fig.~\ref{fig:mag-hist}.

To further validate the resulting photometry, we compare our cluster sample with catalogues from the LEGUS survey for the dwarf galaxies UGC~685 and UGC~1249 \citep{cook_star_2019}, which span a similar range of stellar masses and star-forming dwarf environments to NGC~1569. UGC~685 is a lower-mass, lower-metallicity irregular/spiral dwarf with log(M$\star$/\Msun) = 7.9 and 12+log(O/H) = 8.0, while UGC~1249 is a disturbed irregular dwarf with log(M$\star$/\Msun) = 9.0 and near-solar metallicity (12+log(O/H) = 8.62) \citep{cook_spitzer_2014}. Together, these systems bracket the stellar mass and metallicity of NGC~1569, providing a useful comparison for assessing the consistency of the derived cluster photometry.

\begin{figure*}
    \centering
    \includegraphics[width=0.4\linewidth]{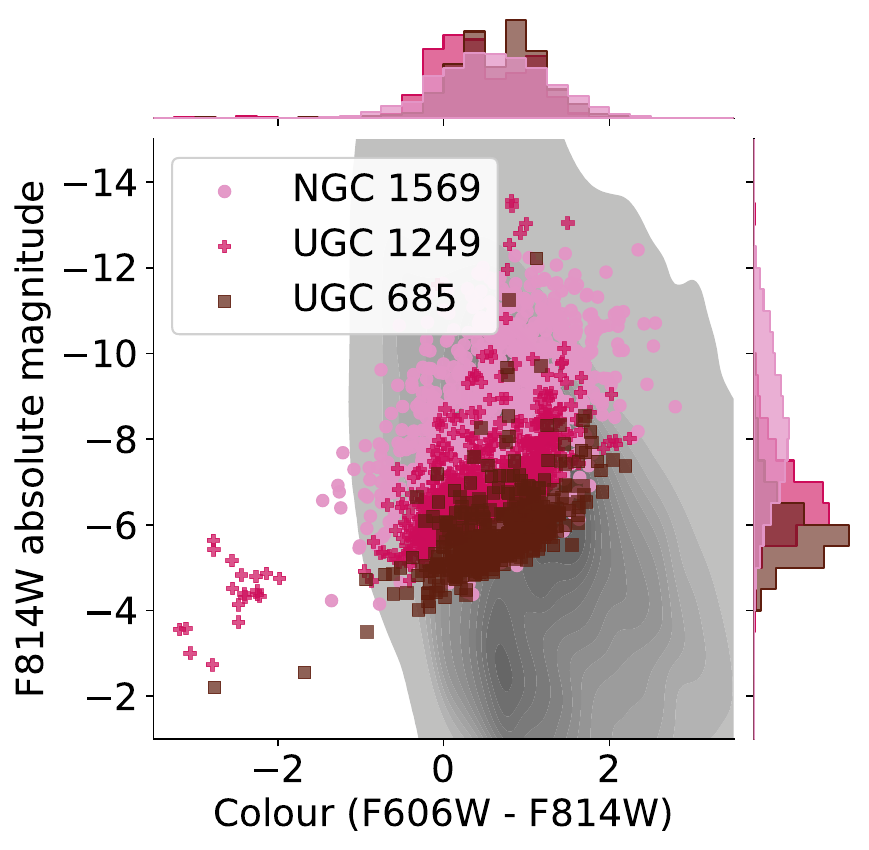}
    \includegraphics[width=0.4\linewidth]{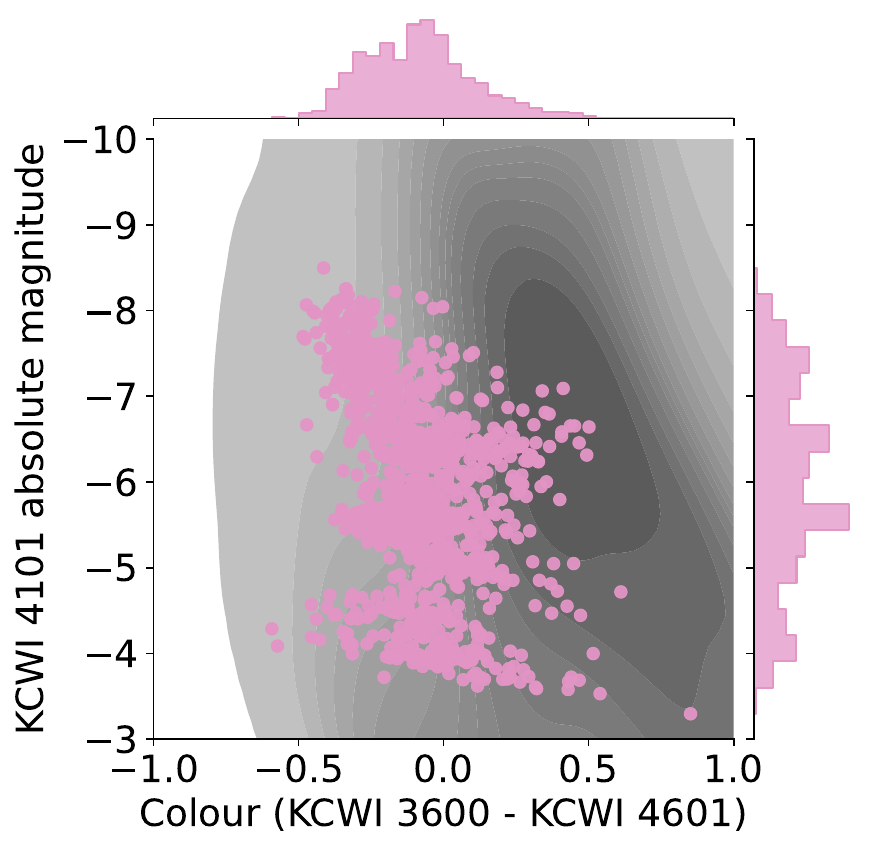}
    \caption{\textbf{Left:} CMD showing the photometry of clusters in NGC~1569 (pink circles) and the LEGUS clusters in UGC~1249 (magenta crosses), and UGC~685 (brown squares). The contours in the background show the density of the simulated cluster library. The vertical axis shows the absolute magnitude in F814W, the horizontal axis shows the colour defined F606W - F814W. The marginal plots are density histogram of each of the catalogues over both axes. \textbf{Right:} CMD comparing the extracted KCWI photometry to the synthetic photometry. The vertical axis shows the magnitude in KCWI 4101, and the horizontal shows a colour defined by KCWI 3600 - KCWI 4601. The pink points show our observed photometry, while the countours show the simulated library. In both panels, the overall colour distribution is comparable between the observed and synthetic photometry.}
    \label{fig:CMDs}
\end{figure*}

As the LEGUS catalogues use Vega magnitudes, we convert our fluxes to Vega magnitudes using Vega zeropoints from \verb|acstools| \citep{acstools}. We then compare the absolute magnitudes of the clusters. A colour-magnitude diagram (CMD) containing the LEGUS clusters, the NGC~1569 clusters, and the \verb|SLUG| generated clusters (see Section~\ref{sec:SimPhot}) can be seen in the left hand panel of Figure \ref{fig:CMDs}. We see that there is overall good agreement in the colours of the clusters, but the clusters in NGC~1569 are overall brigher than the others. Although the brightest clusters in the two LEGUS catalogues are brighter than the brightest clusters in our NGC~1569 list.

A similar CMD for the KCWI photometry is shown in the right panel of Figure \ref{fig:CMDs} (given the custom nature of the KCWI filters, no survey comparison can be made). Since the KCWI data does not have associated Vega magnitude zeropoints, we convert the fluxes to Vega magnitudes using the \verb|cluster_slug| function \verb|convert_photometry|, which finds the Vega zeropoint in each band using the Vega spectrum.

HST photometric uncertainties are computed using the error estimation method given in the SExtractor documentation \citep{SExtractor},

\begin{equation}
    \mathrm{Flux error} = \sqrt{\sum_{i\in A}\left(\sigma_i^2+\dfrac{p_i}{g_i}\right)}
\end{equation}

where $A$ is the set of all pixels in the aperture, $\sigma_i^2$ is the noise variance estimated from the local background, $p_i$ is the background subtracted pixel value, and $g_i$ is the effective detector gain in $e^-/\mathrm{ADU}$.

For the KCWI data, variance cubes are available; however, for our purposes the formal instrumental uncertainties are not the dominant source of error. Instead, the uncertainty is driven by the extraction of cluster photometry from the KCWI data cube (e.g. aperture definition, background subtraction, and spatial blending). We therefore adopt fractional photometric uncertainties drawn from a Gaussian distribution with a mean of 20\% and dispersion of 5\%, which provides a reasonable representation of the expected extraction uncertainties. The impact of different error assumptions on the derived results is explored in Appendix~\ref{app:err_comp}, where we compare fits obtained using alternative uncertainty estimates.

\subsection{Synthetic Photometry}\label{sec:SimPhot}
When determining the physical properties of a star cluster, there is no unique or deterministic mapping between observed photometry and intrinsic parameters. For example, a high flux in a band sensitive to massive stars does not necessarily imply a high cluster mass: while this may hold statistically, stochastic sampling of the IMF means that a low-mass cluster may occasionally host a few massive stars and produce similar luminosities. In addition, photometric measurements are affected by degeneracies between parameters such as age and extinction, both of which can produce faint and red colours. To interpret the photometry, we therefore compare the observations to simple stellar population (SSP) models, which represent coeval stellar populations with a single metallicity while varying parameters such as age, total mass, and extinction. By matching the model photometry to the observed photometry, we can infer the physical properties of the clusters.

Here, we use the \verb|SLUG| (\textbf{S}tochastically \textbf{L}ighting \textbf{U}p \textbf{G}alaxies) software suite \citep{silva_slugstochastically_2012, 2014MNRAS.444.3275D, krumholz_slug_2015} to generate SSPs that stochastically sample the IMF and to peform Bayesian inference on the properties of the observed clusters using the library of SSPs as a reference.
Generating SSPs with \verb|SLUG| requires specifying a number of parameters describing both the stellar population and its environment. The environmental parameters—including the visual extinction $A_V$, the shape of the extinction curve, and the fraction of ionising photons absorbed by hydrogen rather than dust ($\phi$)—determine the effects of dust attenuation and nebular emission in the synthetic clusters. The adopted parameter ranges are motivated by the results of \cite{MagdaPaper}, who analysed the gas and outflows in NGC~1569 using the same KCWI IFU data employed here. We sample $A_V$ from a flat distribution between 0 and 10 mag, adopt the \verb|SLUG| LMC extinction curve (UV–optical from \cite{LMC_uv_opt} and IR from \cite{LMC_ir}), and set the fraction of ionising photons absorbed by hydrogen to $\phi = 0.73$ \citep{1997ApJ...476..144M}. The LMC extinction law is chosen because the metallicity of NGC~1569 is comparable to that of the LMC, making it a reasonable approximation for the dust properties in this system.

The stellar population parameters include the IMF, metallicity, CMF, a choice of stellar evolution tracks, and a time sampling function to choose the age distribution of our clusters. Here we use a Kroupa IMF \citep{Kroupa02} and a simple power-law CMF $dN/dM\propto M^{-2}$, ranging from $M=1\times10^2$\Msun\ to $M=1\times10^8$\Msun. The age distribution of the synthetic clusters follows a power-law, where $dN/dT\propto-1$, ranging from $t=0.1$~Myr to $t=15$~Gyr. The shapes of these functions are not required to be realistic, and are weighted to place more synthetic clusters where stochasticity is more relevant, i.e. at lower masses and younger ages. We adopt the Padova stellar evolution tracks including AGB stars with a metallicity of $Z = 0.2 $ Z$_{\odot}$, compatible with the known metallicity of NGC~1569.

With these parameters we generate 10$^6$ clusters with photometry in the same bandpasses as our observed data, with the addition of a number of JWST/NIRCAM bands: F070W, F090W, F115W, F150W, F200W, F227W, F356W and, F444W, and the HST/UVIS bands: F275W and F438W. These filters are included to assess and quantify to what extent additional data from JWST and HST would allow obtaining better constraints of cluster properties (see Section~\ref{sec:setups}) if these were available. As the KCWI data is flux calibrated to account for the response curve, we define the passthrough as a constant 100\% over the wavelength range of the KCWI bands in our synthetic photometry.

The synthetic photometry generated by \verb|SLUG| is shown in CMDs in Figure \ref{fig:CMDs} as the greyscale shaded contours, along with the observed photometry of NGC~1569 in both figures, as well as photometry from the literature for comparison purposes in the figure on the left (see Section \ref{sec:observed-photometry}). The observed photometry overlaps quite well with the synthetic over most of the colour space, with the exception of the reddest clusters, which may be due to these being affected by additional extinction or differential reddening which is not replicated by our models.

\section{Filter Selection}\label{sec:FilterSelection}
\begin{figure}
    \centering
    \includegraphics[width=\linewidth]{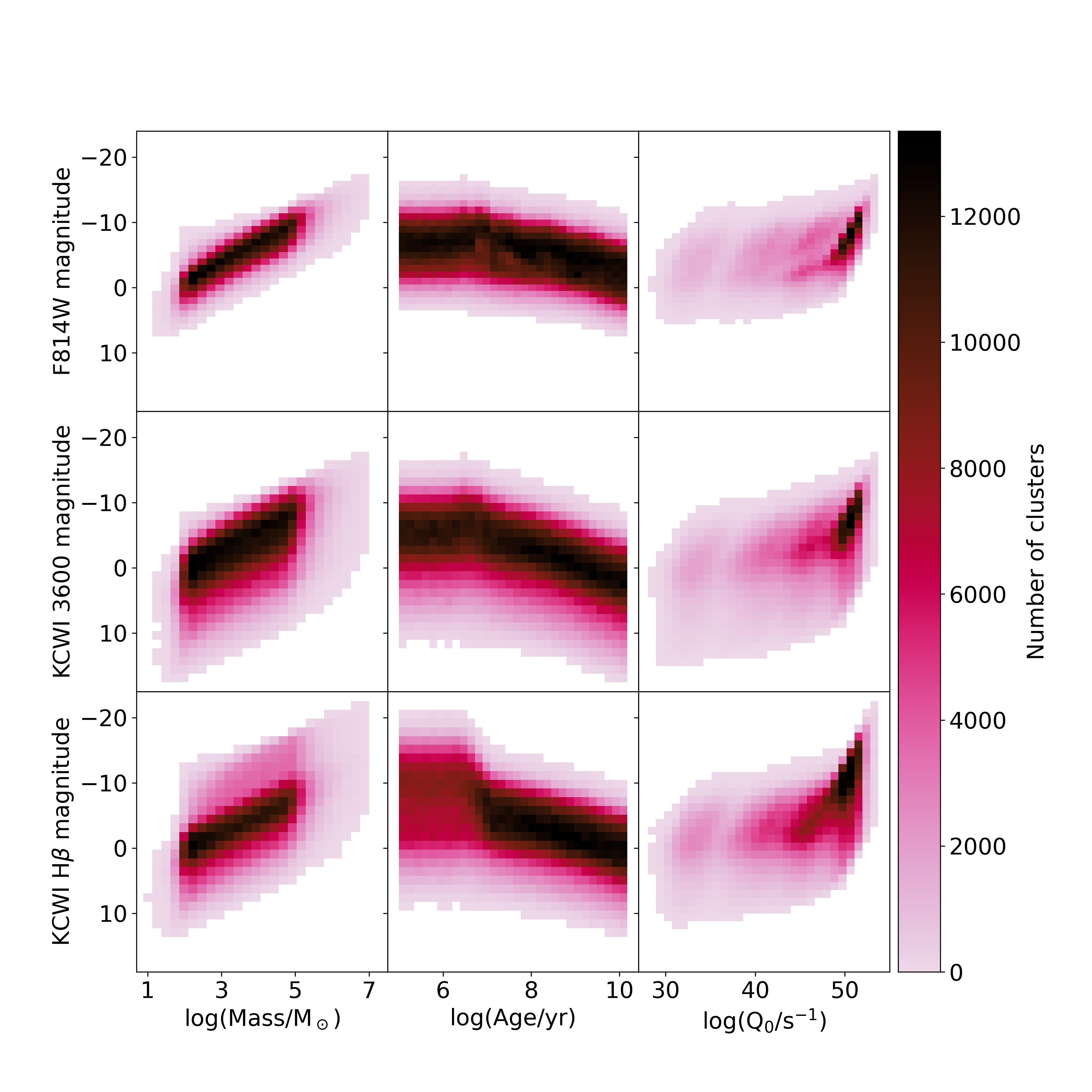}
    \caption{Correlations between synthetic cluster magnitudes and intrinsic properties from SLUG simulations, including nebular emission and extinction. Each panel shows the magnitude in a different band as a function of cluster mass (left), age (middle), and ionising photon flux $Q_{0}$ (right). Broad red bands (F814W) are most sensitive to cluster mass but weakly dependent on age and $Q_{0}$, while narrowband \protect\hbeta\ is strongly age- and $Q_{0}$-sensitive. Intermediate-width blue bands (e.g., KCWI 3600) provide complementary information. These correlations illustrate why broad and intermediate-width bands are sufficient to robustly constrain cluster properties, motivating our final filter selection.}

    \label{fig:mag-prop}
\end{figure}

A critical decision in the process of deriving posterior probability distributions for physical cluster properties (age, mass, extinction, etc.) is which photometric bands to include in the likelihood calculation, since each band contributes different information about the underlying stellar population. Examination of the synthetic photometry (Fig.~\ref{fig:mag-prop}) shows that all filters brighten with increasing cluster mass and fade with age, with the age sensitivity strongest in the narrowband \hbeta\ filter and weakest in the red F814W band. \hbeta\ also correlates most strongly with the ionising photon flux $Q_{0}$, while broad bands such as F814W trace the continuum emission from longer-lived, lower-mass stars, providing robust constraints on stellar mass. These trends illustrate how different bands probe complementary aspects of the cluster population, motivating the use of both broad- and intermediate-width filters in the fitting.
Narrowband filters such as \hbeta\, while highly sensitive to young, massive stars, are excluded from the final fitting set. This choice is motivated by practical considerations: the KCWI pixel size is large enough that nebular emission can be blended between multiple clusters, and in older clusters \hbeta\ can appear in absorption, for which the SLUG stellar atmosphere libraries have limited spectral resolution and therefore may not predict the flux accurately. Focusing on broad- and intermediate-width bands allows us to constrain cluster properties robustly while avoiding potential biases from spectral-resolution limitations or spatial blending.

In the following section we outline the process of selecting the combination of photometric filters used to derive the physical proeprties of the clusters. In Section \ref{sec:mock-observed} we describe the generation of mock observations used for testing purposes. We describe the method of obtaining cluster properties from the photometry in Section \ref{sec:inf-method}, before testing each filter set on the mock observations in Section \ref{sec:mock-test} and on the observed data in Section \ref{sec:obs-test}.

\subsection{Mock-observed catalogue}\label{sec:mock-observed}
In addition to the available filters, we also investigate the potential benefit of including additional bluer and redder photometry from HST and JWST, respectively. To do so, our reference library also includes the HST/UVIS bands F275W and F438W, and the JWST/NIRCam bands F070W, F200W, F227W and F444W. These bands would serve to better constrain the ages of the youngest (bluest) clusters, as well as allowing for better constraints on the lower-mass and/or embedded population with the IR from NIRCam.

To test how well we can recover cluster properties using different filter combinations, we generate an additonal set of 1000 synthetic clusters, in each filter, and treat this set as an observational sample by adding photometric uncertainties. For the synthetic KCWI magnitudes, the uncertainties are computed in the same manner as for the real observations, while the HST and JWST uncertainties are sampled from a Gaussian with $\mu = 0.05\%$ and $\sigma = 0.02\%$.

\subsection{Inference Method}\label{sec:inf-method}
To derive the physical properties - mass, age, and extinction - of the observed clusters, we use the \verb|SLUG| companion Python module \verb|cluster_slug|. Cluster properties are derived via Bayesian inference on the photometry to estimate the physical properties of the observed clusters using the simulated library as a kind of 'training set'. For this we need to provide priors to be used for the Bayesian analysis. We assume the same priors used in \citet{krumholz_slug_2015}.
\begin{equation}\label{eq:priors}
p_{\mathrm{prior}}(\mathbf{x})\propto M^{-1}T^{-0.5}
\end{equation}

Where \textbf{x} is (log~$M$, log~$T$, $A_v$). This prior assumes a cluster mass distribution $dN/dM\propto M^{-2}$, in line with what is expected from observations of young star clusters \citep{bik_clusters_2003, zwart_young_2010, fouesneau2012analyzing, schulz_mass_2015, krumholz_star_2019}. The age distribution is a bit more complex, as below $10^{6.5}$ years cluster dispersion is unlikely to be a factor, while above this age, there is debate on whether the distribution is flat in age ($dN/dT\propto T^0$), or flat in log age ($dN/dT\propto T^{-1}$). A compromise of $dN/dT\propto T^{-0.5}$ is used. We assume a flat prior for the extinction $A_V$. A more detailed motivation of these priors is given in \cite{krumholz_slug_2015}. It is worth noting that the derived cluster properties can be quite sensitive to the choice of priors when the cluster colours are such that the posterior distribution has multiple peaks, although the mass distributon is quite robust for masses from $10^3-10^6$\Msun \citep{krumholz_star_2015}.

\begin{figure}
    \centering
    \includegraphics[width=1\linewidth]{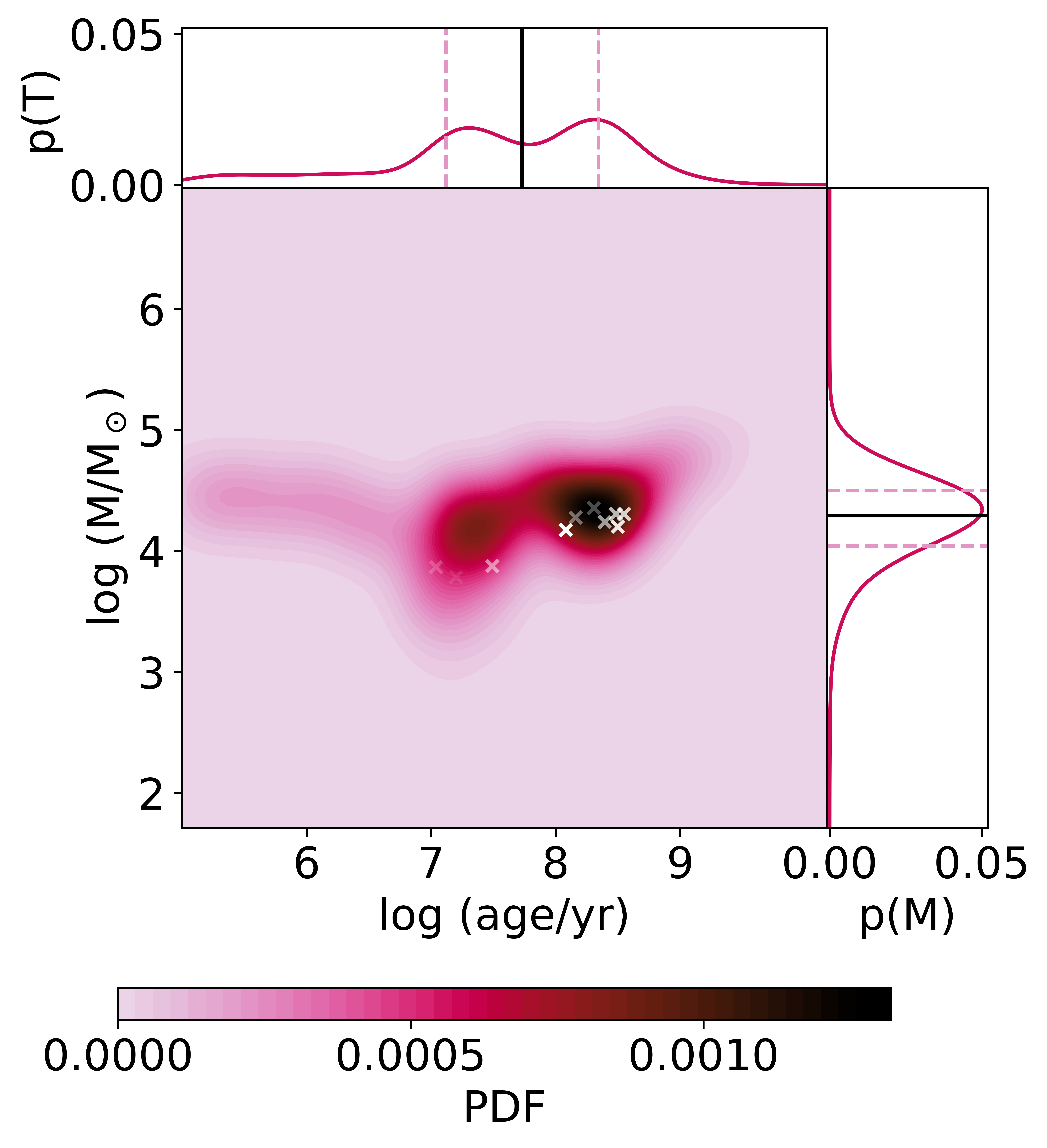}
    \caption{Corner plot showing results of Bayesian inference for one cluster. Each of the marginal plots has a black solid line showing their median and two dotted pink lines for the 25th and 75th percentiles. From this we see that using the median along with these percentiles is a good representation of the distribution, including much of the two separate peaks in the age distribution, while using the peak value would simply set the age at the higher of the two peaks.
    The crosses in the contour plot show the properties of the 10 most photometrically similar clusters in our simulated library, with the most opaque markers being the best matches.}
    \label{fig:cornerplot}
\end{figure}

With these priors, we use \verb|cluster_slug| to obtain posterior probability distributions for the physical properties of the clusters. Each property can be obtained individually or in conjunction with others. For each cluster, we infer the mass and age, and determine the 10 most photometrically similar clusters. 
The cluster properties can be inferred within the 3D space of log~$M$, log~$T$ and $A_v$, yielding a 3D posterior PDF, or marginalised over one or two of these dimensions, yielding 1D or 2D marginal PDFs. In this work we infer both marginal 1D PDFs for each quantity individually, and 2D PDFs for each pair of quantitites. For each cluster we therefore obtain 6 PDFs: 3 1D PDFs when the properties are fit individually, and 3 2D PDFs. In this work, we present results using the simultaneous fit of mass and age, yielding a 2D posterior PDF, which we can then marginalise over each axis to obtain marginal posteriors for age and mass. In principle, cluster extinction could be included as an additional free parameter in the \verb|SLUG| fitting. We tested this by performing fits in which extinction was inferred simultaneously with cluster age and mass. However, the resulting extinction estimates were weakly constrained and did not show clear trends with other cluster properties. Since the inclusion of extinction substantially increases the computational cost of the inference while not materially affecting the derived age–mass distributions, we restrict the analysis presented here to the two-dimensional inference in age and mass. We show an example of a 2D mass-age posterior for one cluster in Fig.~\ref{fig:cornerplot} where the properties of the 10 best photometrically matched synthetic clusters are indicated with crosses. To get the posterior of the full sample when we have obtained 2D PDFs for all our clusters, we normalise each posterior PDF and take the mean.

\verb|cluster_slug| returns probability distributions for each parameter, not a single ``best-fit'' solution. In this context, the median is a natural point esimator because it represents the value that divides the posterior mass in half and is therefore less sensitive to skewness, multi-modality, or long tails than a maximum-likelihood or best-matching solution. This is particularly important for low- and intermediate mass clusters, where stochastic sampling of the IMF may produce highly non-Gaussian posteriors.

\subsection{Assessing filter setups}\label{sec:setups}
\subsubsection{Testing on mock observations}\label{sec:mock-test}
We use \verb|cluster_slug| together with the synthetic library of $10^6$ clusters, to derive PDFs for age and mass, and compare the inferred vs. actual properties for each of the mock observed clusters. As will be described below, with this we can quantify how well the inferred properties matches the true values given a choice of used photometric filter, providing motivation for which filter set to choose to derive the most reliable constraints. 

\begin{figure}
    \centering
    \includegraphics[width=0.9\linewidth]{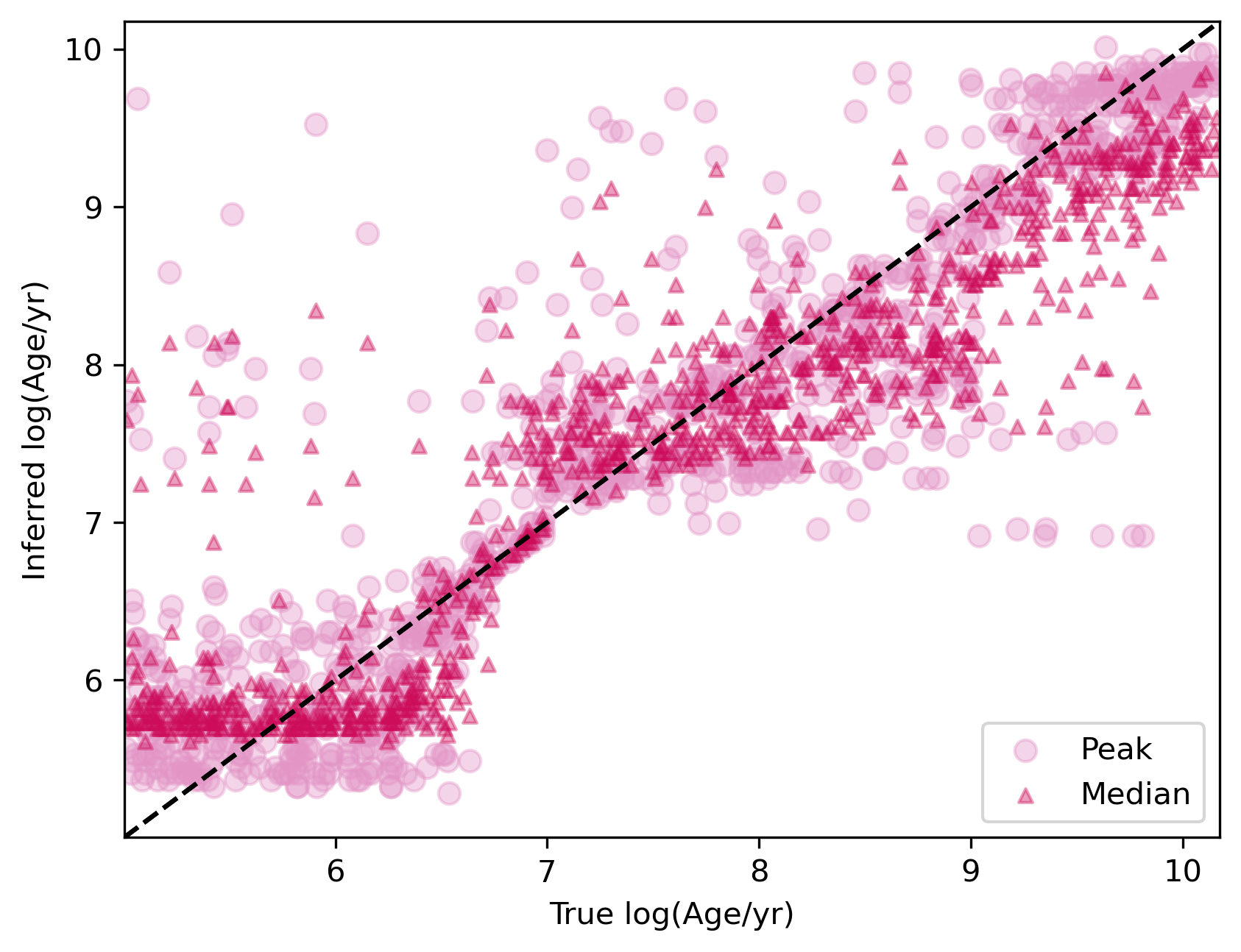}
    \caption{Scatter plot comparing the inferred age of the synthetic cluster population to the ages from the generated catalogue using all available photometric filters (i.e., HST/ACS and KCWI broad- and narrowband). The black dashed line shows the 1:1 line. Magenta triangles show points determined as a median of the distribution, and pink circles show the results when using the peak. We see that although the age is generally well constrained, their is a grouping of the inferred age at $\sim10^6$~yrs, whether the peak or the median is used.}
    \label{fig:inf-true-medvpeak}
\end{figure}

A first comparison using the two available HST filter as well as KCWI broad- and narrow-bands (i.e., all available filters), and using the peak value of the PDF and the median of the distributions to infer cluster properties, is shown in Fig.~\ref{fig:inf-true-medvpeak}.
We see that there is a bunching of ages just below $10^6$ yrs. Looking at the central column of Fig.~\ref{fig:mag-prop} we see that below these ages, there is no change in the magnitude in any of our bands until this age, so as the likelihood function is flat, the posterior PDF is simply the prior. Hence, the values here are simply the median/peak of the prior in this age range. We also see a few catastrophic misses, where the true age is $\lesssim10^{6}$ but the inferred age is far older. These are likely to be young clusters where there is no massive star, meaning they lack nebular emission. Although there are examples of this in the reference library as well, these are a minority and as such \verb|SLUG| will assume that in the absence of nebular emission, the cluster is old. This is a correct assumption a majority of the time, but there are some cases where this causes these catastrophic misses. For these reasons, we cannot put strong constraints on these young ages with our current data.

We then run the inference for multiple combinations of our filters, to showcase the value of including the different bands in retrieving true cluster parameters. We test five different combinations of the available filters:

\begin{itemize}
    \item HST/ACS only (F606W and F814W)
    \item KCWI (broad-band filters) only
    \item KCWI broad- and narrow-band filters
    \item No narrowband: HST and KCWI broad-bands only
    \item All: HST and KCWI broad- and narrow-bands
\end{itemize}

We also test the value of including additional filters by considering:

\begin{itemize}
    \item JWST/NIRCam (F0707W, F200W, F227W, and F444W)
    \item HST/UVIS (F275W and F438W)
    \item All available filters + NIRCam and UVIS
\end{itemize}

\begin{figure*}
    \centering
    \includegraphics[width=\linewidth]{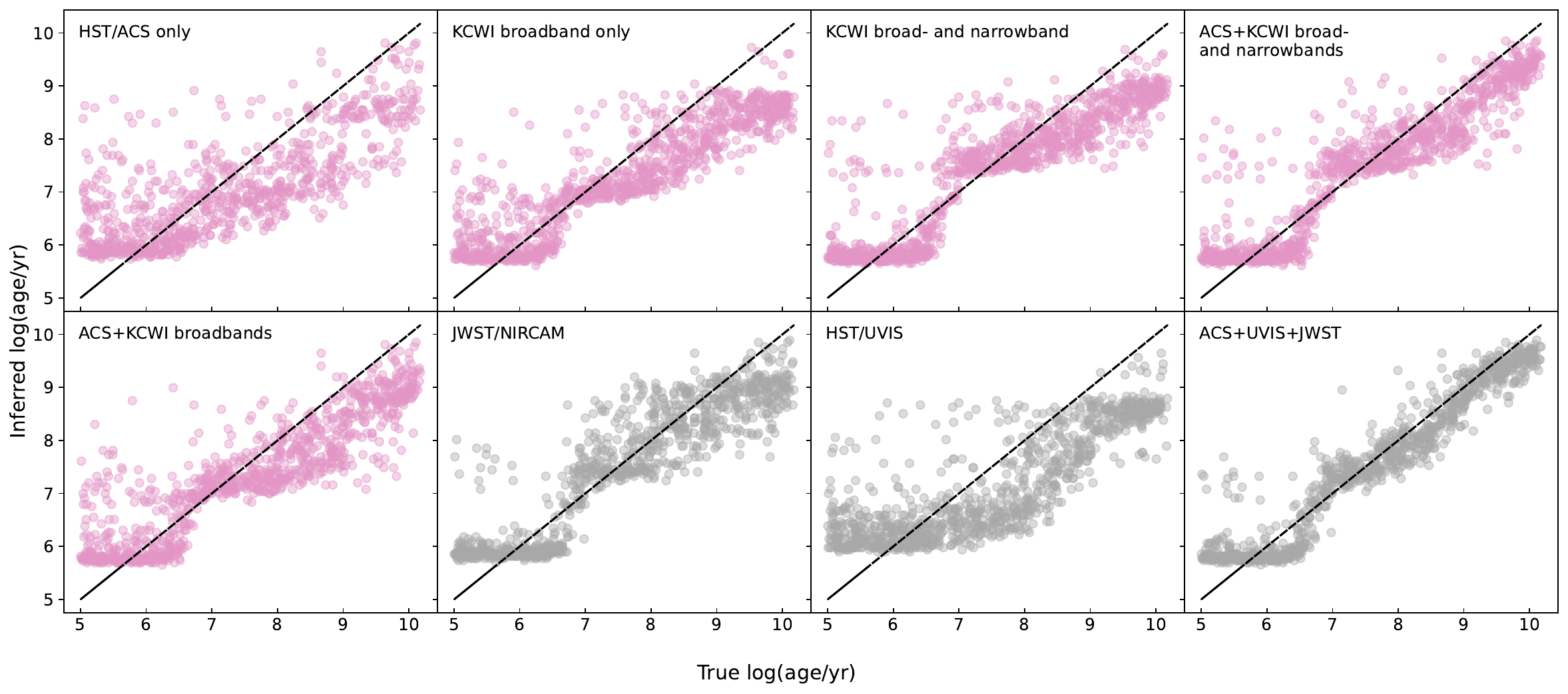}
    \caption{Inferred vs true age for the 1000 simulated observations. From left to right, the top row shows: only ACS F606W and F814W, only the KCWI broadbands, KCWI broadbands and narrowbands, ACS along with KCWI broad- and narrowbands. The bottom row shows: ACS with KCWI broadbands, JWST NIRCAM bands, HST UVIS bands, ACS, UVIS and, NIRCAM bands. The the final three are shown in grey to emphasise that we do not have access to these bands in our observed data. Similar to Fig.~\protect\ref{fig:inf-true-medvpeak} there is a bunching of ages at $\sim10^6$~yrs for all filter sets. We also see that the inclusion of bands covering wider wavelength ranges serves to greatly reduce the scatter.}
    \label{fig:inf-true}
\end{figure*}

Fig.~\ref{fig:inf-true} shows the results of all filter combinations as a comparison between the Bayesian inference parameters to the true parameter values of the synthetic cluster library \verb|SLUG|. 

It is very clear from the figure that increasing the spectral coverage of the data greatly reduces the scatter between the true properties and those inferred. To quantify this reduced scatter we define a distance in physical parameter space,
\begin{equation}
    D_p = \sqrt{\sum_i(p_{i,\mathrm{true}}-p_{i,\mathrm{inf}})^2}
\end{equation}
where $p_{i,\mathrm{inf}}$ is the 50th percentile value (median) in the 1D marginal PDFs and the true value parameter $p_i$ and $p_{i,\mathrm{true}}$ is the value read from the library for the same cluster. $D_p$ is evaluated for each cluster in the library, and we then use the mean as a representation of the scatter for each filter set.
These numbers can then be used to quantify how well each set of filters constrains the properties of the clusters. The mean $D_p$ for each set of filters is presented in Table~\ref{tab:filter-sets}.

\subsubsection{Testing on real data}\label{sec:obs-test}
We run the inference for the same filter sets on the real data, excluding, of course, those setups containing unavailable UVIS or NIRCAM bands. As a measure for the performance of a given combination of filters we monitor the proportion of cluster that are not fit. Poorly fit clusters are those clusters that have no peak in their posterior PDF after Bayesian inference, defined by setting the minimum height of the peak over the baseline to be over 0.01 in the 1D marginal PDF for both mass and age. This does not occur when using synthetic data, but must be considered when it comes to choosing which filter set to use, so we have run the inference on each of the filter sets and record the number of clusters that have no peaks in their PDFs. We then also want to eliminate any filter sets that too strongly assign clusters an age of $<10^6$~yrs. For this reason we also quantify the number of very young clusters once poorly fit clusters are removed. Both of these are displayed in Table~\ref{tab:filter-sets}

\begin{table}
    \centering
    \begin{tabular}{|c|c|c|c|}
    \hline
        \textbf{Filter set} & \textbf{$D_p$} [dex] & \textbf{Poor fits} & \textbf{Very young}\\
        \hline
         HST/ACS+HST/UVIS+JWST & 0.39 & N/A & N/A\\
        \hline
         ACS + KCWI broad and narrow & 0.48 & 62\% & 59\%\\
         \hline
         JWST only & 0.51 & N/A & N/A\\
         \hline
         KCWI broad- and narrowband & 0.57 & 3\% & 81\%\\
        \hline
         ACS + KCWI broadbands & 0.59 & 30\% & 19\%\\
         \hline
         KCWI broadband & 0.69 & 1\% & 2\%\\
        \hline
         HST/ACS only & 0.75 & 6\% & 18\%\\
         \hline
         HST/UVIS only & 0.88 & N/A & N/A\\
         \hline
    \end{tabular}
    \caption{The parameter distance $D_P$, the number of poorly fit clusters, and the number of clusters assigned very young ages ($<10^6$ yrs) for the different filter sets presented. Those sets that have N/A in the Poor Fits and Age column do not have real observed data and as such can not have any poorly fit clusters.}
    \label{tab:filter-sets}
\end{table}

These results clearly demonstrate the added value of including both IR and UV photometry for reducing scatter and alleviating the age degeneracy. We also find that, although the inclusion of narrowband filters leads to tighter constraints on cluster properties, nearly half of the clusters cannot be successfully fit with this combination of filters. This is likely due to the relatively large pixel size of the KCWI data, which can cause the nebular emission to be spatially blended. As a result, the narrowband flux associated with a given cluster may be significantly contaminated by ionising radiation from neighbouring clusters, leading to the assignment of nebular luminosities to the wrong source.

\subsection{Quality Flags}
Prior to presenting the results of the Bayesian inference in the next section, we determine we determine quality flags for the inferred clusters in order to assess how well the observed photometry is reproduced by the synthetic cluster library. Due to the uncertain nature of the photometry from the IFU data, it is important to identify which clusters are well represented by the models and therefore have reliably constrained physical properties. To do so we compute an error-normalised photometric distance between each observed cluster and its potential matches in the synthetic library, using only the broad-band magnitudes. Following \citet{krumholz_slug_2015}, this distance is defined as
\begin{equation}
    D_M = \sqrt{\dfrac{1}{N}\sum_i\dfrac{(M_{i,\mathrm{obs}}-M_{i,j})^2}{\Delta M_{i,\mathrm{obs}}}}.
\end{equation}

where $N$ is the number of photometric filters, $M_{i,\mathrm{obs}}$ is the magnitude of the observed cluster in filter $i$, $\Delta M_{i,\mathrm{obs}}$ is the uncertainty on that magnitude, and $M_{i,j}$ is the magnitude of the $j$th synthetic cluster in the same filter.
For each observed cluster we compute $D_M$ relative to every synthetic cluster in the library, and use the resulting distribution of distances to assign a quality flag. Clusters with at least five library matches satisfying $D_M \leq 1$ , corresponding to matches within the $1\sigma$ photometric uncertainty, are assigned a quality flag $Q = 0$. Clusters that do not meet this criterion but have at least five matches with $D_M \leq 2$ are assigned $Q = 1$. All remaining clusters are assigned $Q = 2$, indicating that the observed photometry is poorly reproduced by the model library. In the analysis that follows we focus primarily on clusters with $Q = 0$ and $Q = 1$, while clusters with $Q = 2$ are excluded from further consideration. After inference and the removal of poorly fit clusters, this procedure yields 432 clusters with $Q = 0$, 455 with $Q = 1$, and 228 with $Q = 2$.

\section{Analysis}\label{sec:Analysis}
In what follows, we present the results of the Bayesian inference of the properties of the observed clusters. Specifically, in Section \ref{sec:bayes-results} we compare the cluster ages to the star formation history of the galaxy, and in Section \ref{sec:spatial_analysis} we analyse the spatial distirbution of cluster masses and ages across NGC~1569. Comparing the inferred cluster ages to the galaxy’s star formation history provides an important consistency check between independent tracers of recent and past star formation episodes. Star clusters form preferentially during episodes of elevated star formation, and thus the distribution of cluster ages should broadly reflect the temporal variation of the star formation rate. Examining the spatial distribution of cluster masses and ages allows us to investigate how star formation has propagated across the galaxy, e.g. spatial gradients in age as a tracer for feedback effects or dynalical processes, and variations in cluster mass as a tracer for environmental effects such as gas density, pressure, or feedback from previous star formation episodes.

\begin{figure*}
    \centering
    \includegraphics[width=0.47\linewidth]{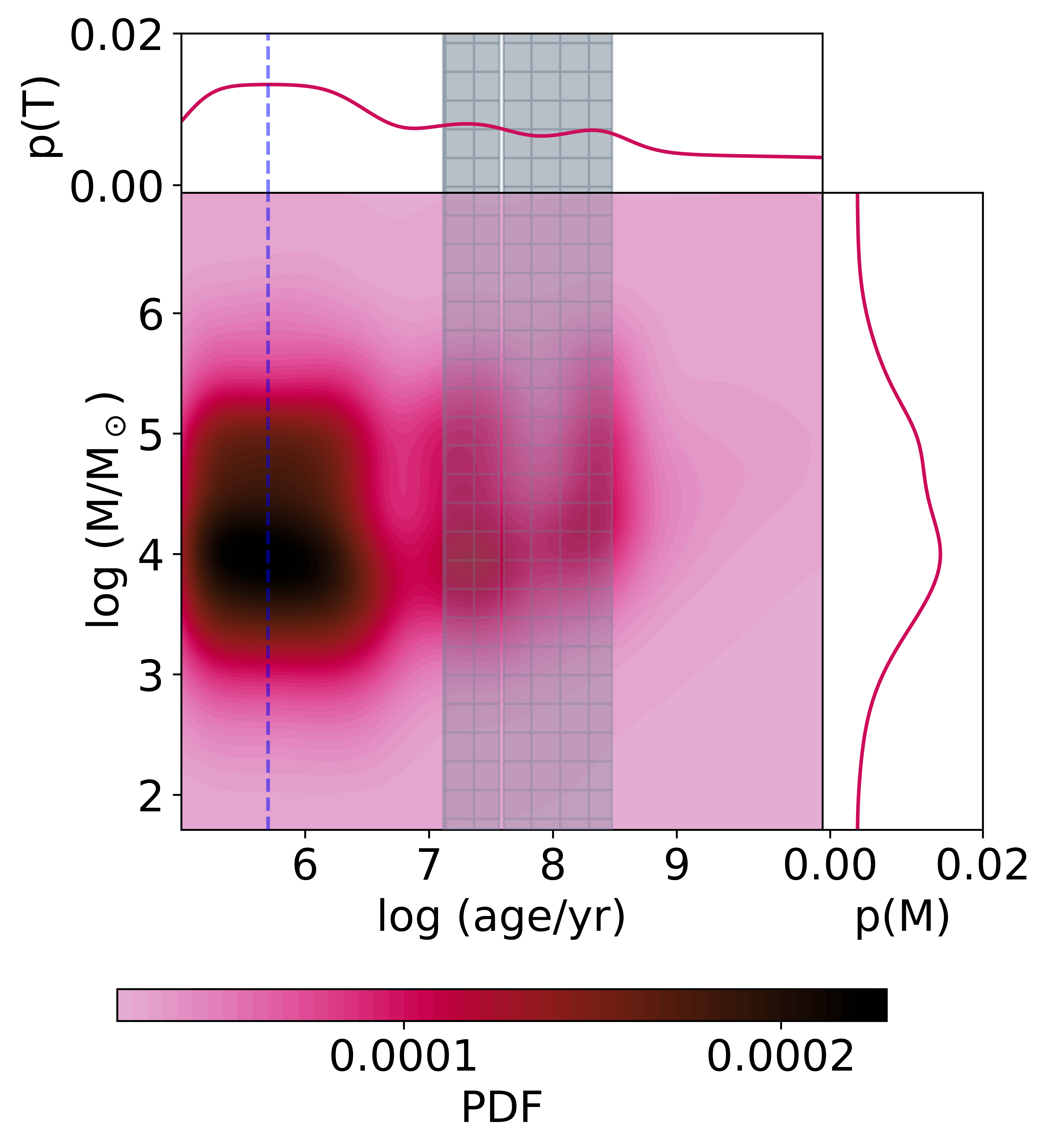}
    \includegraphics[width=0.525\linewidth]{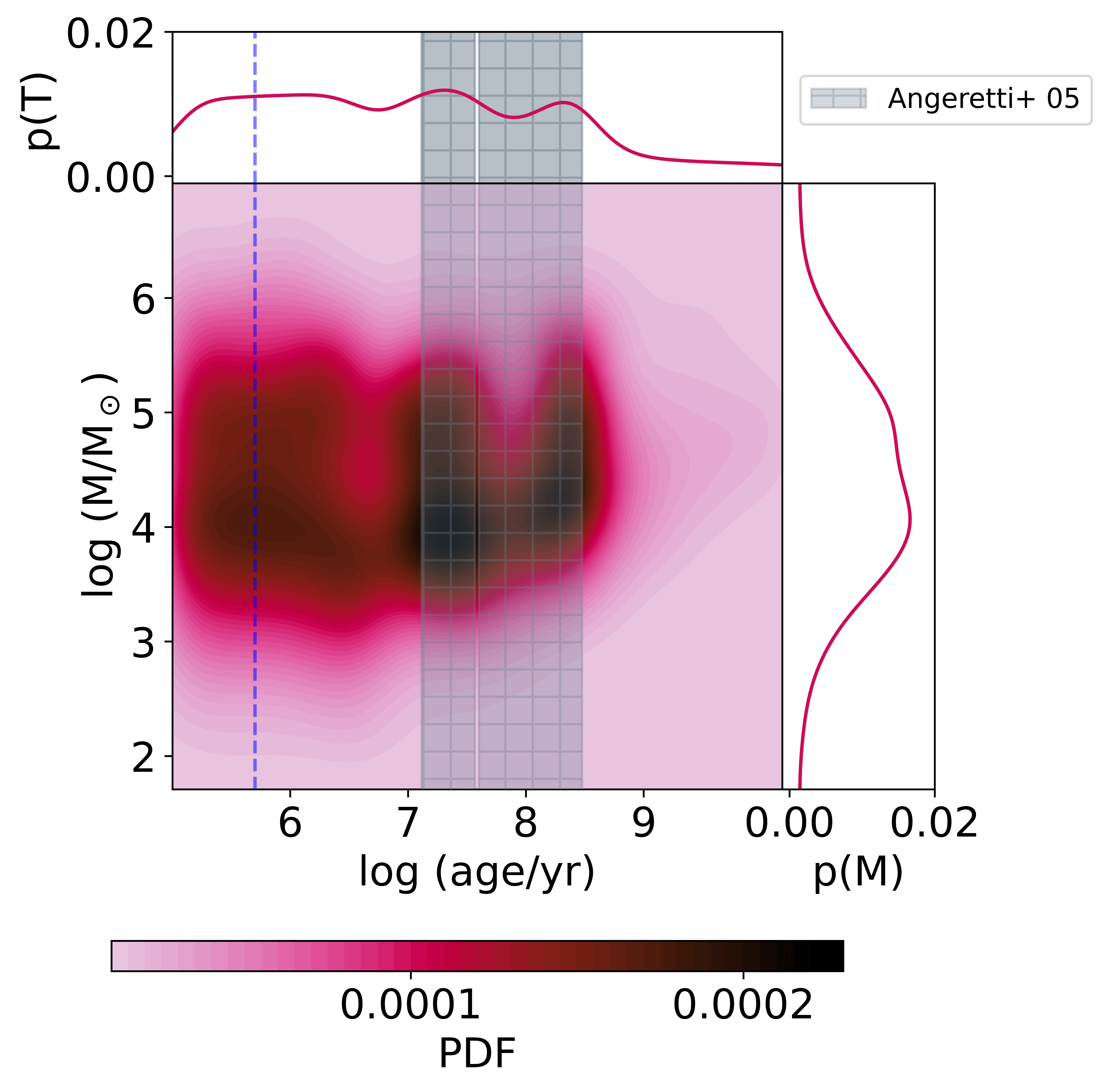}
    \caption{Contour plots showing the results of Bayesian inference. The vertical axis shows the inferred mass and horizontal shows inferred age. The marginalised 1D posterior PDFs are shown in the margins. The shaded regions show the two star formation episodes reported in \protect\cite{angeretti_complex_2005}, along with a blue dashed line identifying the centre of a peak at young ages. The plot on the left shows the full dataset, while for the plot on the right we have removed clusters where the age distribution peaks below 10$^{6}$ years to highlight the other two peaks. In either case we clearly see multiple age peaks, at $10^{5.7}$~yrs, $10^{5.7}$~yrs and, $10^{5.7}$~yrs, with the older two lining up with the results of \protect\cite{angeretti_complex_2005}.}
    \label{fig:results-noNb}
\end{figure*}

\subsection{Properties of observed clusters}\label{sec:bayes-results}

Fig.~\ref{fig:results-noNb} shows the results of the Bayesian inference of age and mass using the ACS filters and the broad KCWI bands, with marginalised distributions in the margins. Due to the artificial bunching of clusters at ages $\lesssim10^6$ yr (see Section \ref{sec:setups}), we apply an additional age cut in the right-hand panel, including only clusters with inferred ages $>10^6$ yr. This allows variations across the age distribution to be more clearly seen while reducing the influence of potentially spurious very young ages. The resulting distribution appears multi-modal, suggesting three main periods of cluster formation. We identify a young component centred at $10^{5.7}$~yrs ($\sim0.5$ Myr), an intermediate population at $10^{7.3}$~yr ($\sim21$~Myr) and an older episode at $10^{8.3}$~yr ($\sim200$~Myr). The number of clusters associated with the youngest peak may be overestimated due to the known pile-up of solutions at very young ages; however, since this feature extends beyond $10^{6}$ yr, we retain it as a possible recent burst of star formation.

Previous studies of the star formation history of NGC~1569 have also identified multiple episodes of enhanced star formation. \citealt{angeretti_complex_2005} use HST photometry and the synthetic CMD technique applied to resolved stars to identify two recent major episodes of star formation occurring between $1.3\times 10^{7}-3.7\times10^{7}$ yr and $4\times 10^{7}-1.5\times10^{8}$ yr. These time intervals are broadly consistent with the two older peaks in our cluster age distribution at $\sim 2\times 10^{7}$ and $\sim 2\times 10^{8}$ yr. While the CMD approach used in \citep{angeretti_complex_2005} and the stochastic sampling Bayesian inference method used here probe different tracers and are subject to different systematic uncertainties, the qualitative consistency between the two results supports a picture in which NGC~1569 has experienced multiple recent episodes of enhanced star formation over the past few hundred Myr. Although our analysis suggests somewhat more pronounced structure in the age distribution, our inferred peaks are also broadly consistent with \citealt{anders_star_2004}, who use multi-band HST photometry and evolutionary synthesis modelling of cluster spectral energy distributions, finding that the majority of clusters formed during a strong recent burst beginning $\sim 25$~Myr ago, with a smaller population of older clusters extending to ages of order $\sim 100$~Myr.

\begin{figure}
    \centering
    \includegraphics[width=\linewidth]{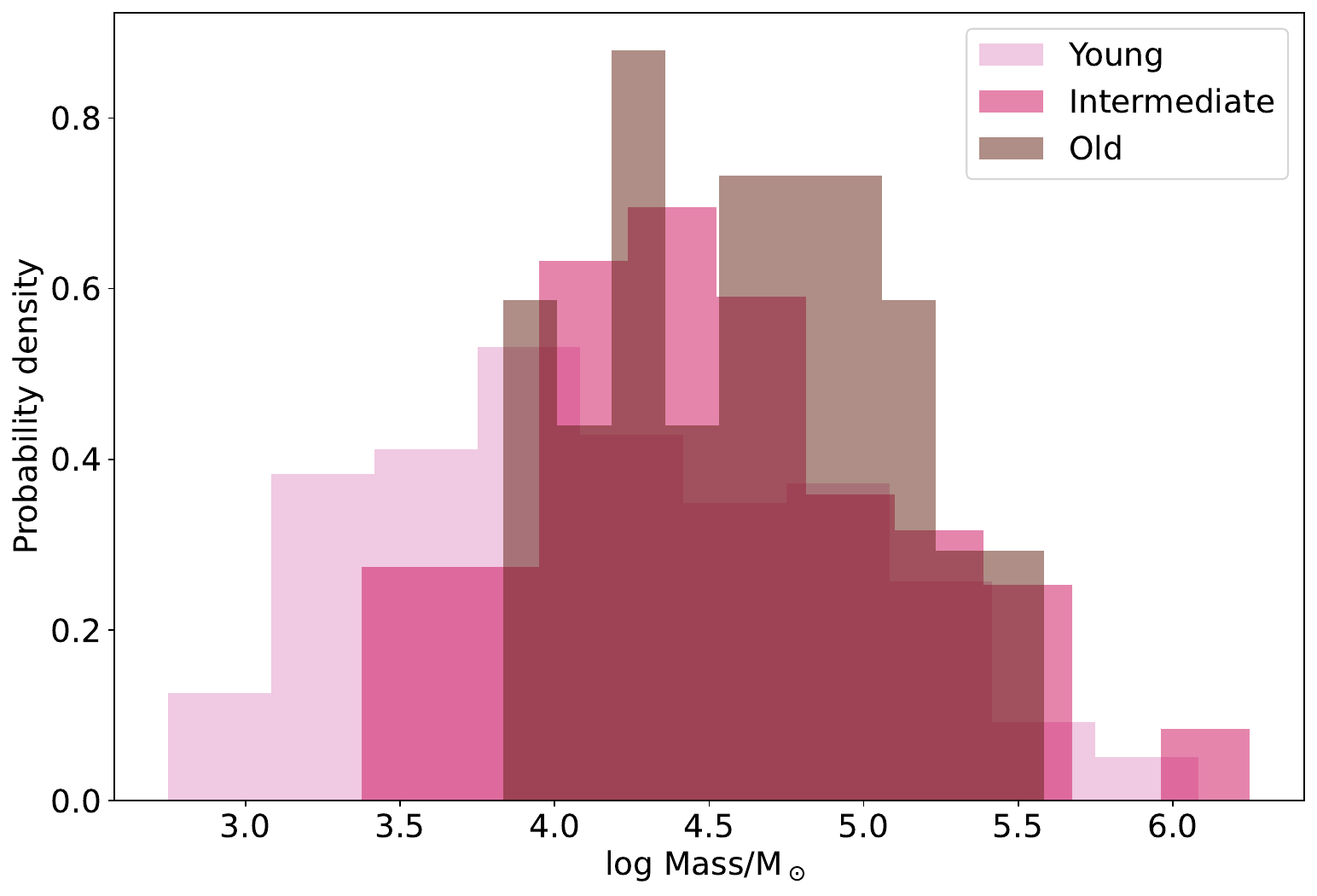}
    \caption{A histogram comparing the cluster masses in the three identified bursts of star formation. The young, intermediate, and old, shown here in pink, magenta, and brown, respectively. Although the lack of low-mass, old clusters is likely due to observational effects, there is a notable difference in the peak of the mass distributions.}
    \label{fig:mass-hist}
\end{figure}

To investigate potential variations in cluster properties over the lifetime of NGC~1569, we divide the clusters into arbitrary age bins roughly corresponding to the identified star formation episodes: young clusters with ages $<10^{6.5}$~yr, intermediate clusters between $10^7-10^{7.5}$~yrs, and old clusters between $10^8-10^{8.5}$~yrs. The cluster mass distributions in these bins are shown in Fig.~\ref{fig:mass-hist}, revealing apparent differences between star formation episodes. To assess whether these differences are statistically significant, we perform Kolmogorov–Smirnov (KS) tests, which evaluate the probability that two samples are drawn from the same underlying distribution. Comparing the young and intermediate clusters yields $p = 3.9\times10^{-8}$, the intermediate and old clusters give $p = 0.42$, and the young and old clusters give $p = 1.26\times10^{-5}$. It is important to emphasise that the KS test does not account for observational limitations, and some apparent differences may therefore be influenced by selection effects. For example, the scarcity of low-mass, old clusters largely reflects their faintness and our limited detection sensitivity. Nevertheless, other differences, such as the location of the mass distribution peaks, suggest that the cluster mass function may genuinely vary between star formation episodes. 

Qualitatively, Fig.\ref{fig:mass-hist} shows that the old population has the highest mean cluster mass, the young population the lowest, and the intermediate is in between these two. Taken together, these results provide tentative evidence that the star formation history of NGC~1569 is not only temporally structured but may also involve variations in the typical masses of clusters formed during different episodes where more recent episodes form progressively less massive clusters. A more detailed interpretation of these trends, accounting for selection effects and cluster disruption, is deferred to Section~\ref{sec:Discussion}.

\subsection{Spatial analysis}\label{sec:spatial_analysis}
In addition to studying how cluster properties relate to one another, we investigate if they vary spatially across the galaxy. To this end, we make spatial maps, overplotting the locations of clusters. We first examine the connection between position of the clusters and their mass.
\begin{figure}
    \centering
    \includegraphics[width=1\linewidth]{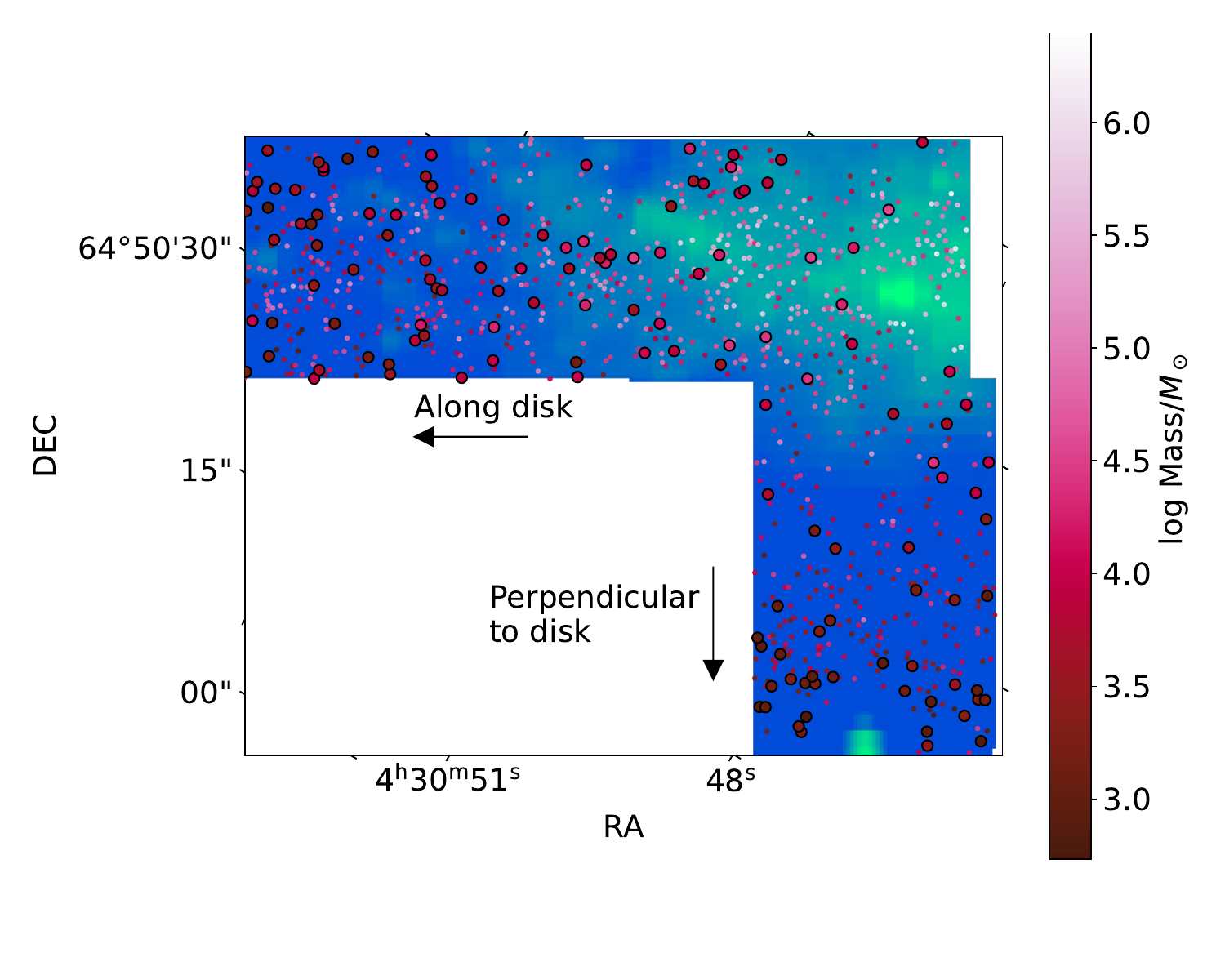}
    \caption{A flattened KCWI image with the locations of clusters marked by points. The outlined points are those clusters that have $Q = 0$ 0, the others have $Q = 1$. The points are coloured by the inferred mass of the clusters in $M_\odot$.}
    \label{fig:Mass-map}
\end{figure}
\begin{figure*}
    \centering
    \includegraphics[width=\linewidth]{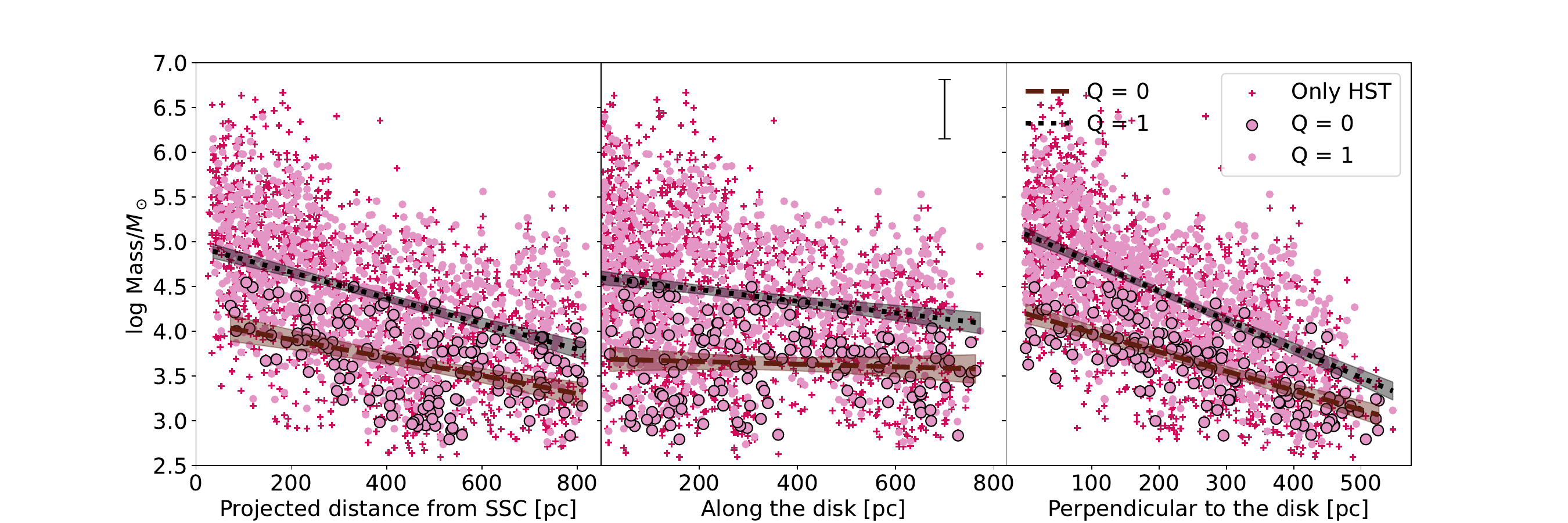}
    \caption{The cluster mass plotted over the distance from SSC~B. The magenta crosses show the mass inferred using only the ACS/HST photometry, the pink points show the masses inferred by also including the KCWI broad bands. The points with outlines show clusters with $Q = 0$, while those without have $Q = 1$. The brown dashed line shows a linear fit to the KCWI data with $Q = 1$, and the black dotted line is a fit to those with $Q = 0$. The average errorbars are shown in the top right of the central frame, both lines have the 95\% confidence interval shaded. We see a clear negative gradient across all three panels, with the slope being steepest when moving perpendicular to the disk, suggesting a change in the trunaction mass $M_c$ of the CMF.}
    \label{fig:mass-dist-maps-both}
\end{figure*}

In Fig.~\ref{fig:Mass-map}, the clusters are plotted over a KCWI integrated-light map and coloured by mass. In Fig.~\ref{fig:mass-dist-maps-both}, we show cluster mass as a function of projected distance from the galactic centre, taken to be the location of SSC~B. We consider the azimuthal distance as well as distances along axes parallel and perpendicular to the disk. A clear gradient in cluster mass is apparent, most pronounced perpendicular to the disk, along the galaxy’s outflow. Distances are projected and not corrected for inclination, which would introduce additional uncertainties

To test whether the observed trends could be driven by observational effects, we consider that clusters near the galaxy centre are more likely to have their fluxes overestimated due to elevated background levels in the KCWI data, and faint clusters may be undetected. To mitigate these effects, we repeat the analysis using only the HST photometry, which employs smaller apertures and is less sensitive to background contamination. We quantify the gradient by performing unweighted linear fits to each dataset along the three distance measures. The resulting slopes are listed in Table~\ref{tab:slope-tab}. For example, using HST-only data, the slope of cluster mass versus azimuthal distance and along the disk is $-1\times 10^{-3}$ pc$^{-1}$, and perpendicular to the disk is $-4\times 10^{-3}$ pc$^{-1}$. Similar slopes are obtained when considering only clusters with quality flags $Q=0$ or $Q=1$. The consistently steeper slope perpendicular to the disk indicates that the mass gradient along this direction is robust and not solely an observational artefact

Even considering potential incompleteness, such as undetected low-mass clusters in bright central regions, there is a clear paucity of high-mass clusters in the outskirts. We also note a systematic difference between $Q=0$ and $Q=1$ clusters: $Q=0$ clusters generally have slightly lower masses. This may result from higher-mass clusters being physically larger, with flux less likely to be fully captured in a single pixel, and preferentially located in denser, higher-background regions, which can increase measurement uncertainty.

\begin{table}
    \centering
    \begin{tabular}{|c|l|l|l|}
    \hline
         \textbf{Dataset}& \textbf{Distance} & \textbf{Along disk}& \textbf{Perp. to disk}\\
         \hline
         HST only& $-1\pm0.2$& $-1\pm0.09$& $-4\pm0.1$\\
         \hline
         $Q = 0$& $-1\pm0.2$& $-1\pm0.2$& $-2\pm0.2$\\
         \hline
         $Q = 1$& $-1\pm0.1$& $-0.7\pm0.1$& $-3\pm0.2$\\
         \hline
    \end{tabular}
    \caption{The fitted slopes for log mass over the azimuthal distance, and over each axis of the disk, for all three shown datasets. All slopes are in units of $10^{-3}$dex~pc$^{-1}$.}
    \label{tab:slope-tab}
\end{table}

\begin{figure}
    \centering
    \includegraphics[width=\linewidth]{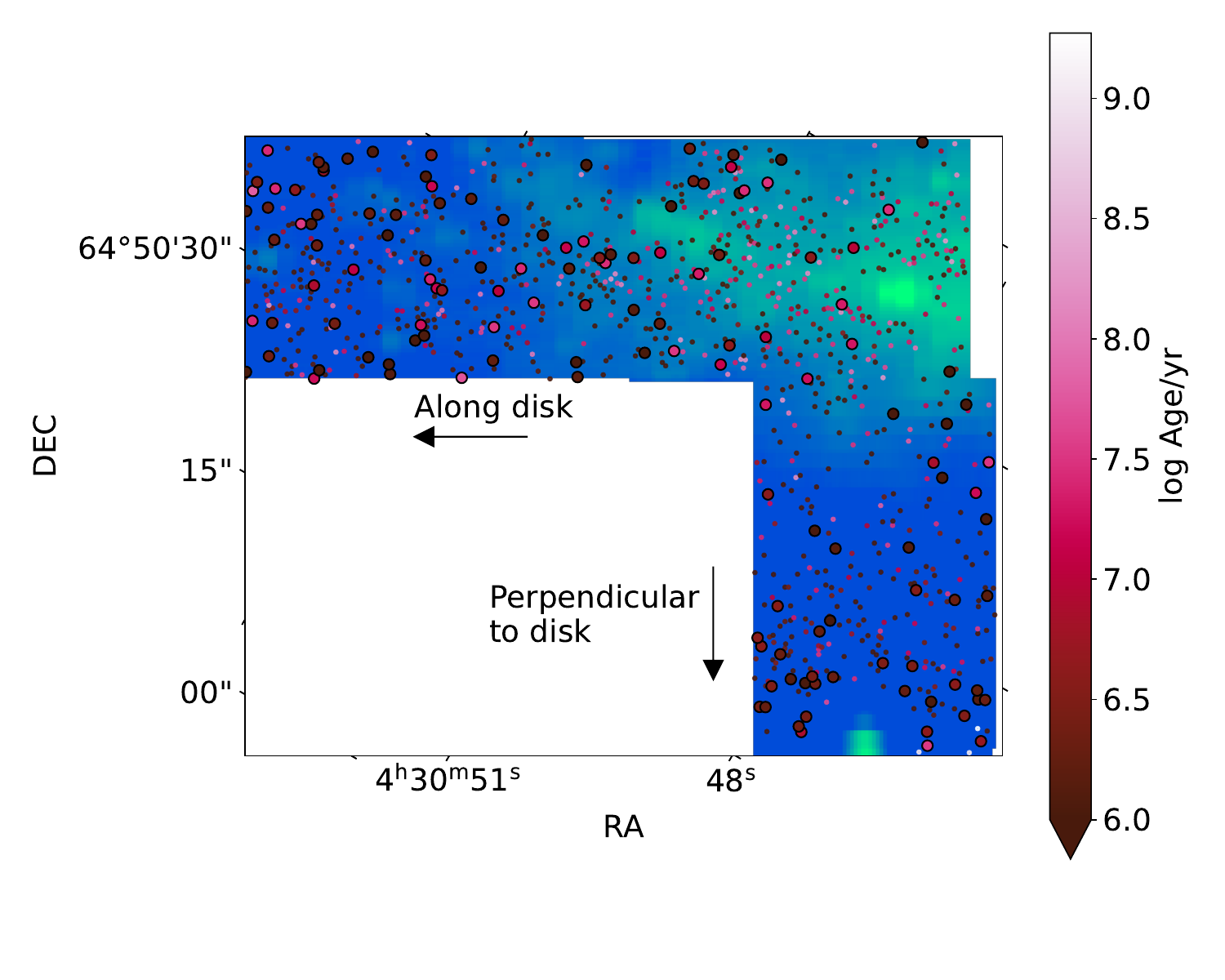}
    \caption{A flattened KCWI image with the locations of clusters marked by points. The outlined points are those clusters that have $Q = 0$, the others have $Q = 1$. The points are coloured by the inferred age of the cluster.}
    \label{fig:Age-map}
\end{figure}

\begin{figure*}
    \centering
    \includegraphics[width=\linewidth]{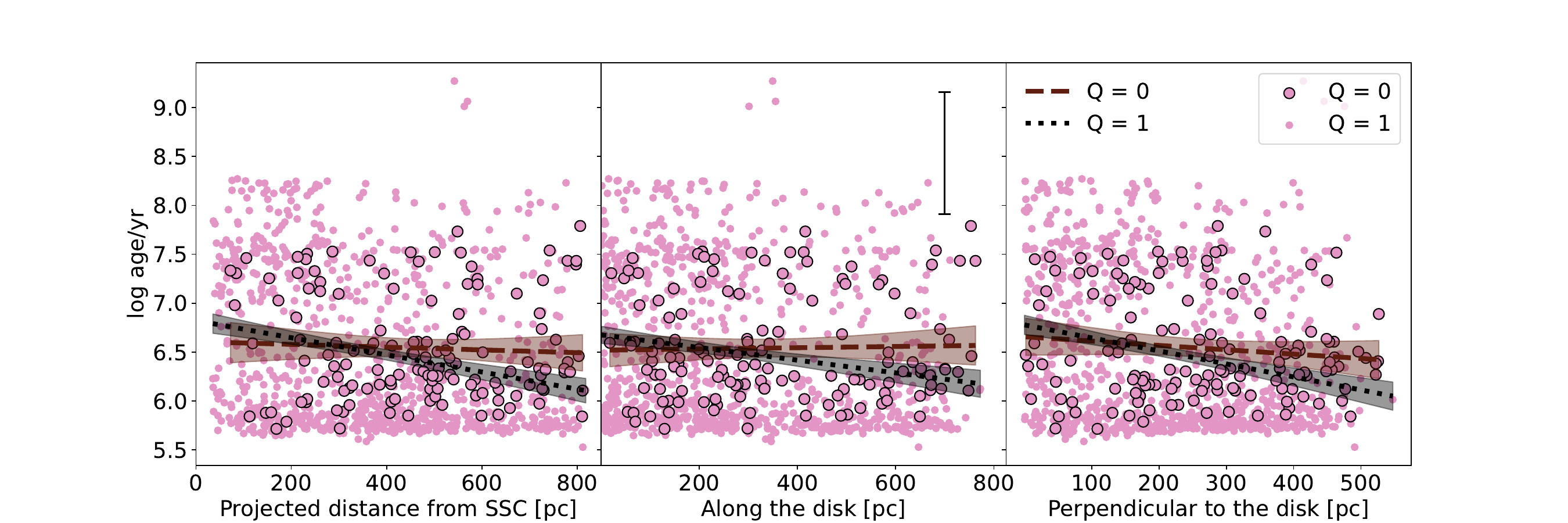}
    \caption{The cluster age plotted over the distance from SSC~B. The points with outlines show clusters with $Q = 0$, while those without have $Q = 1$. We see no clear trend with age over the distance.}
    \label{fig:age-dist-maps}
\end{figure*}

\begin{figure*}
    \centering
    \includegraphics[width=\linewidth]{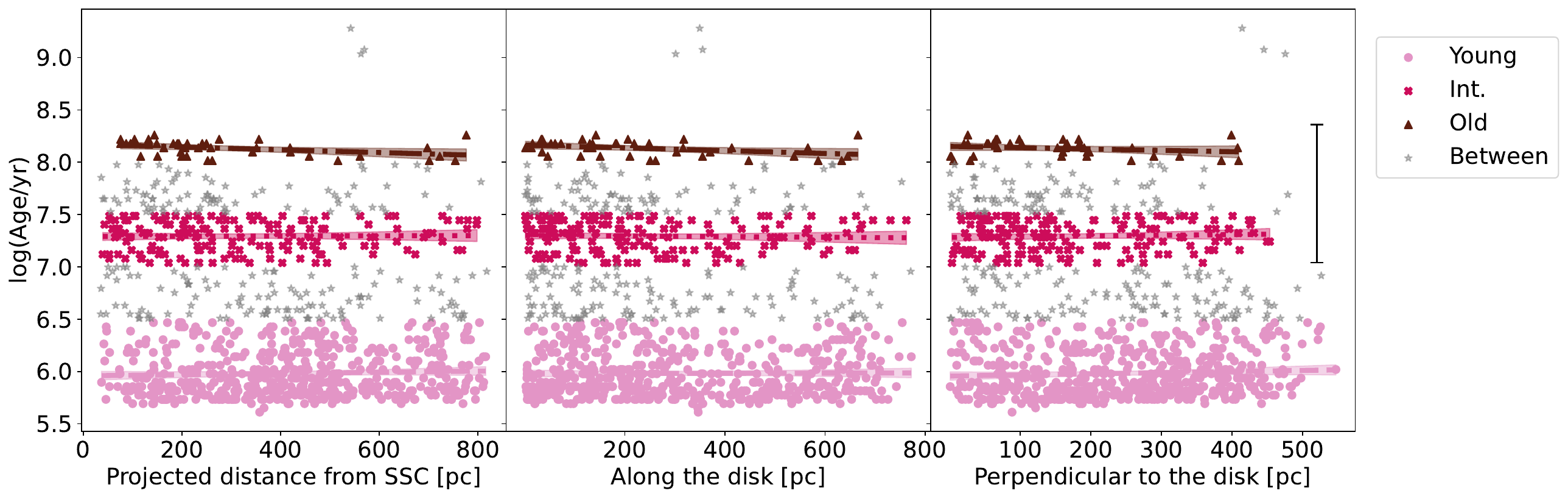}
    \caption{Cluster ages plotted over distance from SSC~B, binned by star formation episode. The pink points show those points in the youngest bin, the magenta crosses those in the intermediate bin and the brown triangles those in the oldest bin. The gray points show those clusters that are not associated with one of the three main star formation episodes identified. The pink dashed line shows a linear fit to the first bin, the red dotted line a fit to the intermediate, and the brown dash-dotted line the oldest. There is a slight negative slope in the oldest age bin, potentially suggesting star formation moving outwards, but note also the large uncertainties on the age determination.}
    \label{fig:age-dist-maps-binned}
\end{figure*}

A similar analysis is performed for the derived cluster ages. The map is shown in Fig.~\ref{fig:Age-map}, where the clusters are now coloured by their age. No obvious variations with galactocentric distance are identified from Fig.~\ref{fig:age-dist-maps}. Further, we also investigate whether there is spatial variation within each star formation episode identified in Section \ref{sec:bayes-results}. To do this we examine their spatial dependence separately. This is shown in Fig.~\ref{fig:age-dist-maps-binned}, where we fit lines to each star formation episode (age bin) separately. The fitted slopes of these lines are reported in Table~\ref{tab:sf-slopes}.

\begin{table}
    \centering
    \begin{tabular}{|c|c|c|c|}
    \hline
        \textbf{Episode} & \textbf{Distance} & \textbf{Along disk} & \textbf{Perp. to disk}\\
        \hline
         Young & $6\pm5$ & $0.8\pm4$ & $10\pm7$\\
         \hline
         Int. & $1.2\pm6$ & $-3\pm6$ & $7\pm9$\\
         \hline
         Old. & $-13\pm6$ & $-13\pm6$ & $-13\pm10$\\
         \hline
    \end{tabular}
    \caption{The fitted slopes for the azimuthal distance, and over each axis, for all three star formation episodes. Slopes are in units of $10^{-5}$dex~pc$^{-1}$}
    \label{tab:sf-slopes}
\end{table}
We find a shallow negative gradient within each age bin, particularly for the oldest episode, potentially suggesting star formation moving from the inside out. It is, however, important to note the large uncertainties on the inferred age and of the linear fits.

\subsection{Correlations with gas properties}
We now investigate the presence of correlations between the physical properties of the clusters and those of the surrounding gas. For this, we use KCWI gas property maps produced for \cite{MagdaPaper}.

\begin{figure*}
    \centering
    \includegraphics[width=0.5\linewidth]{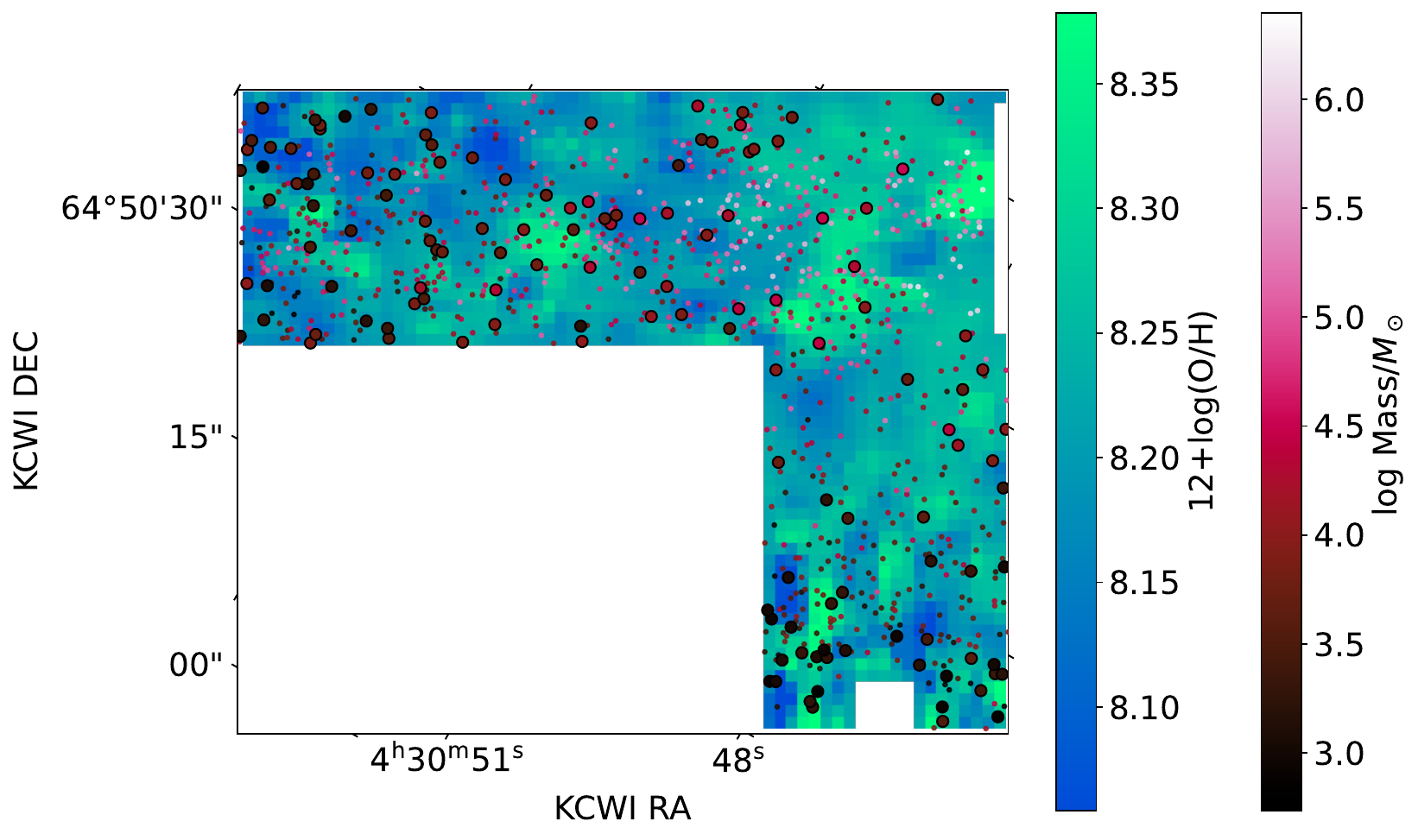}
    \includegraphics[width=0.4\linewidth]{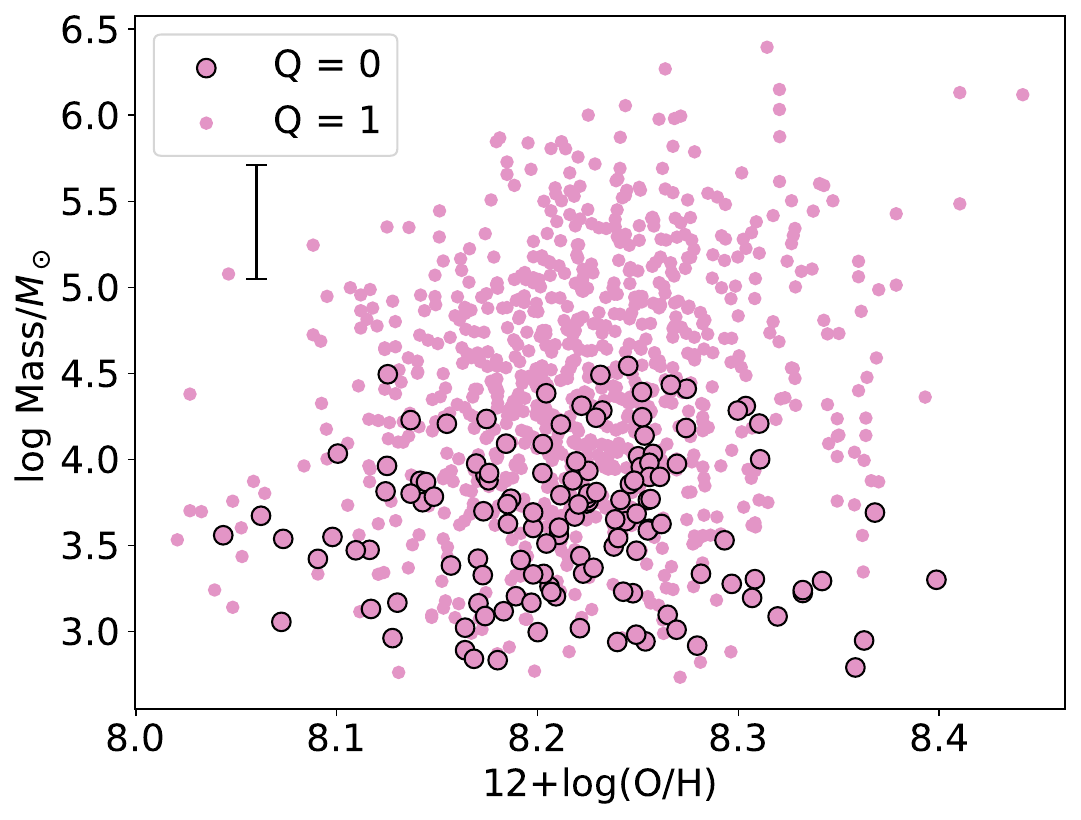}
    \caption{\\
    \textbf{Left: }A map showing the 12+log(O/H) metallicity derived by \protect\citet{MagdaPaper} overlayed by points showing stellar clusters. The points are coloured by their stellar mass.\\
    \textbf{Right: }Scatter plot showing the mass of clusters plotted over their local metallicity. The outlined points show clusters with flag $Q = 0$ while those without an outline have flag $Q = 1$. Although there is no clear linear trend, we see that the highest mass clusters, tend to be in regions of higher metallicity.}
    \label{fig:mass-metals}
\end{figure*}

First, we compare gas-phase metallicity and cluster mass by extracting 12+log(O/H) values from the pixels the clusters are located in. We show this comparison in Fig~\ref{fig:mass-metals}, where the left panel shows a metallicity map, with the location of clusters marked and coloured by their mass, and the right panel shows cluster mass against metallicity.

While the scatter is large, we qualitatively find that the highest mass clusters tend to be in regions of higher metallicity. Quantifying the correlation with the Pearson correlation coefficient (PCC) between the cluster mass and metallicity, we obtain PCC = 0.18, and a p-value for the null hypothesis of $2.6\times10^{-8}$, meaning it is very likely that the mass of the clusters is correlated with the metallicity of the surrounding gas.

\begin{figure*}
    \centering
    \includegraphics[width=0.5\linewidth]{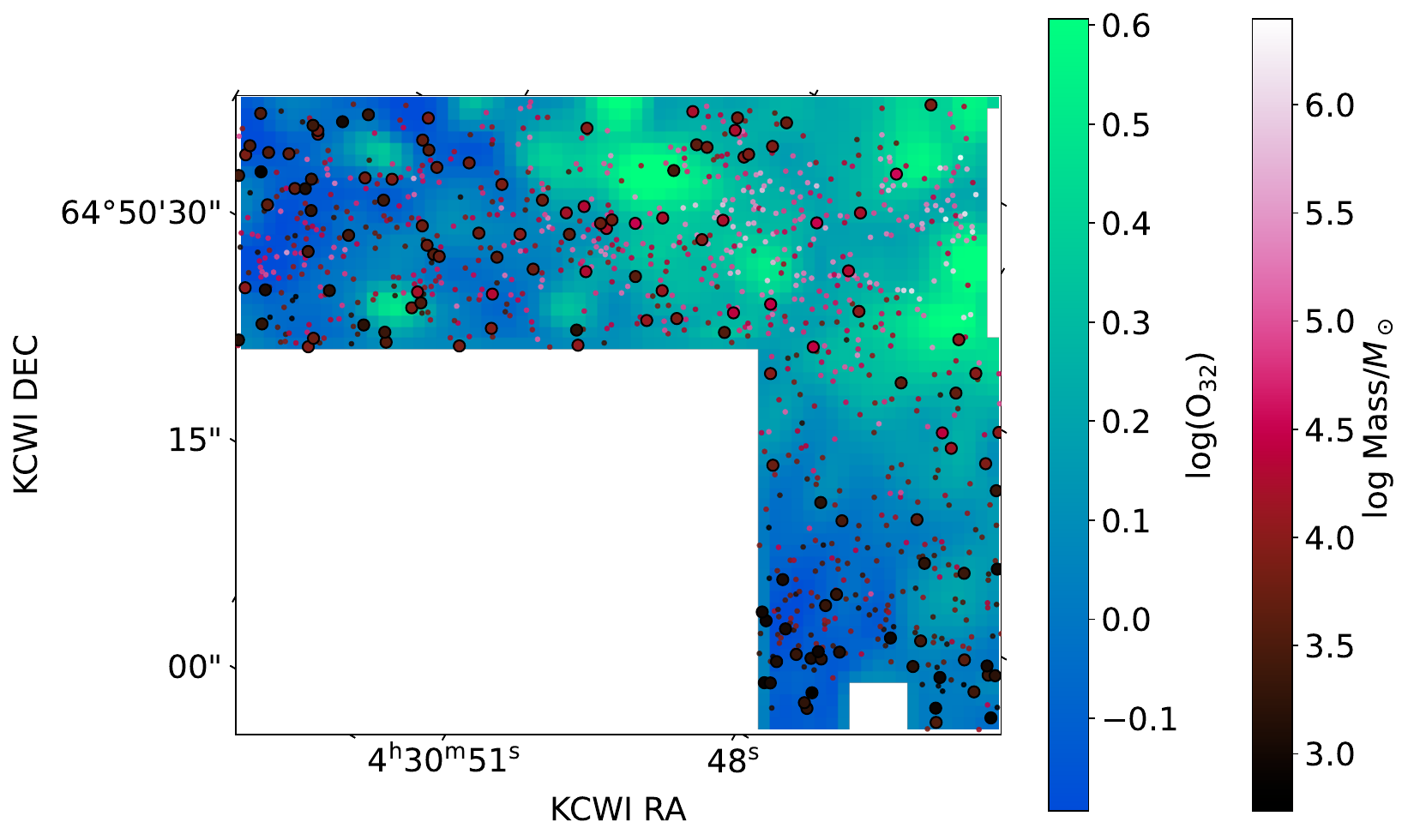}
    \includegraphics[width=0.4\linewidth]{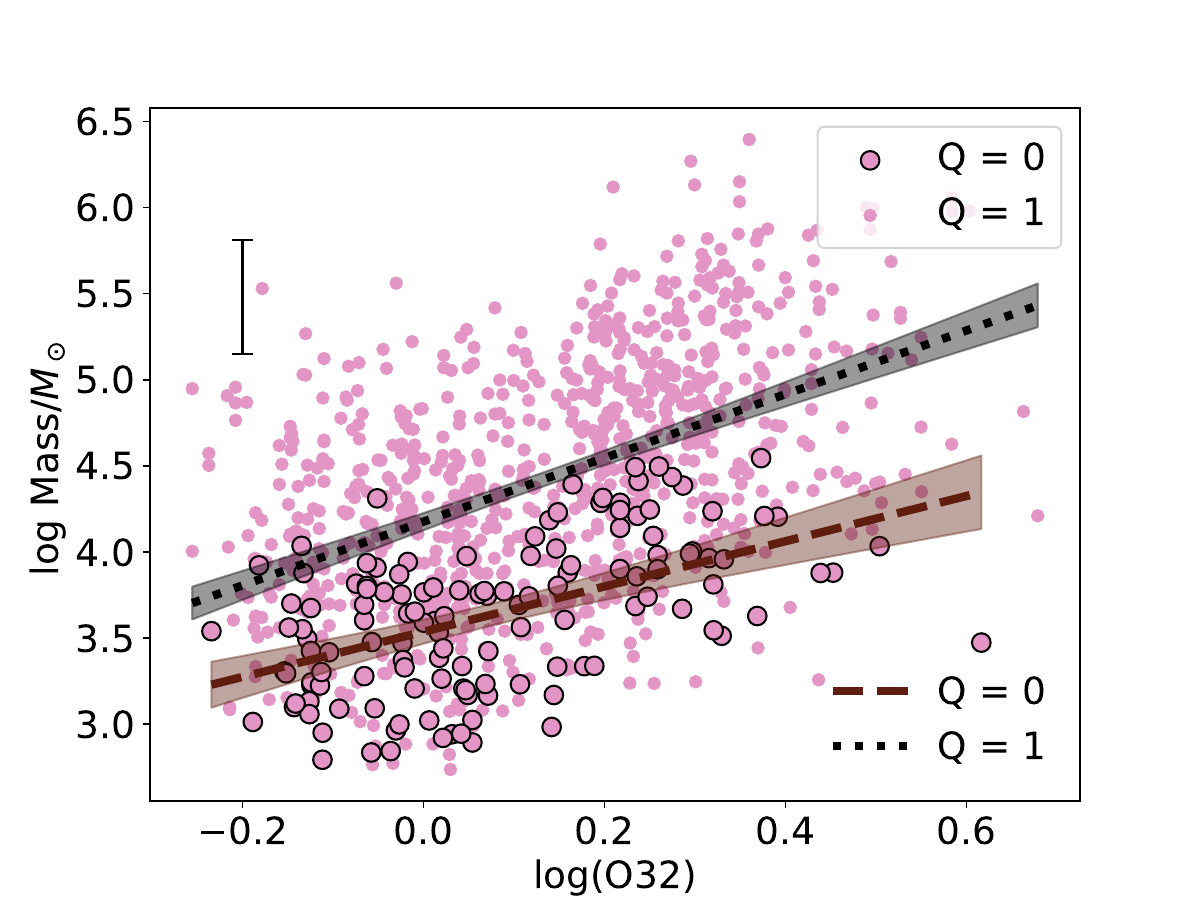}
    \caption{\\
    \textbf{Left: }As Fig.~\ref{fig:mass-metals} but using log(O32) from \protect\citet{MagdaPaper} in place of the 12+log(O/H).\\
    \textbf{Right: }Scatter plot showing the mass over the log(O32). The outlined points show clusters with $Q = 0$ while those without an outline have $Q = 1$. The outlined points show clusters with $Q = 0$ while those without an outline have $Q = 1$. The brown dashed line shows an unweighted linear fit to the $Q = 0$ points and the black dotted line a fit to $Q = 1$. There is a clear positive gradient, with higher mass clusters residing in areas of high ionisation.}
    \label{fig:mass-O32}
\end{figure*}

Next, looking now at the ionisation state of the gas, we use the [\oIII]$\lambda$5007 and [\oII]$\lambda$3727 maps provided by \citet{MagdaPaper} to make a map of the log([\oIII]$\lambda$5007/[\oII]$\lambda$3727), log(O32), to serve as a proxy of the ionisation state of the gas. We can then repeat the mapping process as before, using the log(O32) in place of the metallicity (Fig.~\ref{fig:mass-O32}). 

In these figures there is a clear relationship between cluster mass and ionisation state. Doing an unweighted linear fit to each set of points ($Q = 0$ and $Q = 1$), we get the $Q = 0$ slope $1\pm0.2$ and the $Q = 1$ slope $2\pm0.1$.

\begin{figure*}
    \centering
    \includegraphics[width=0.5\linewidth]{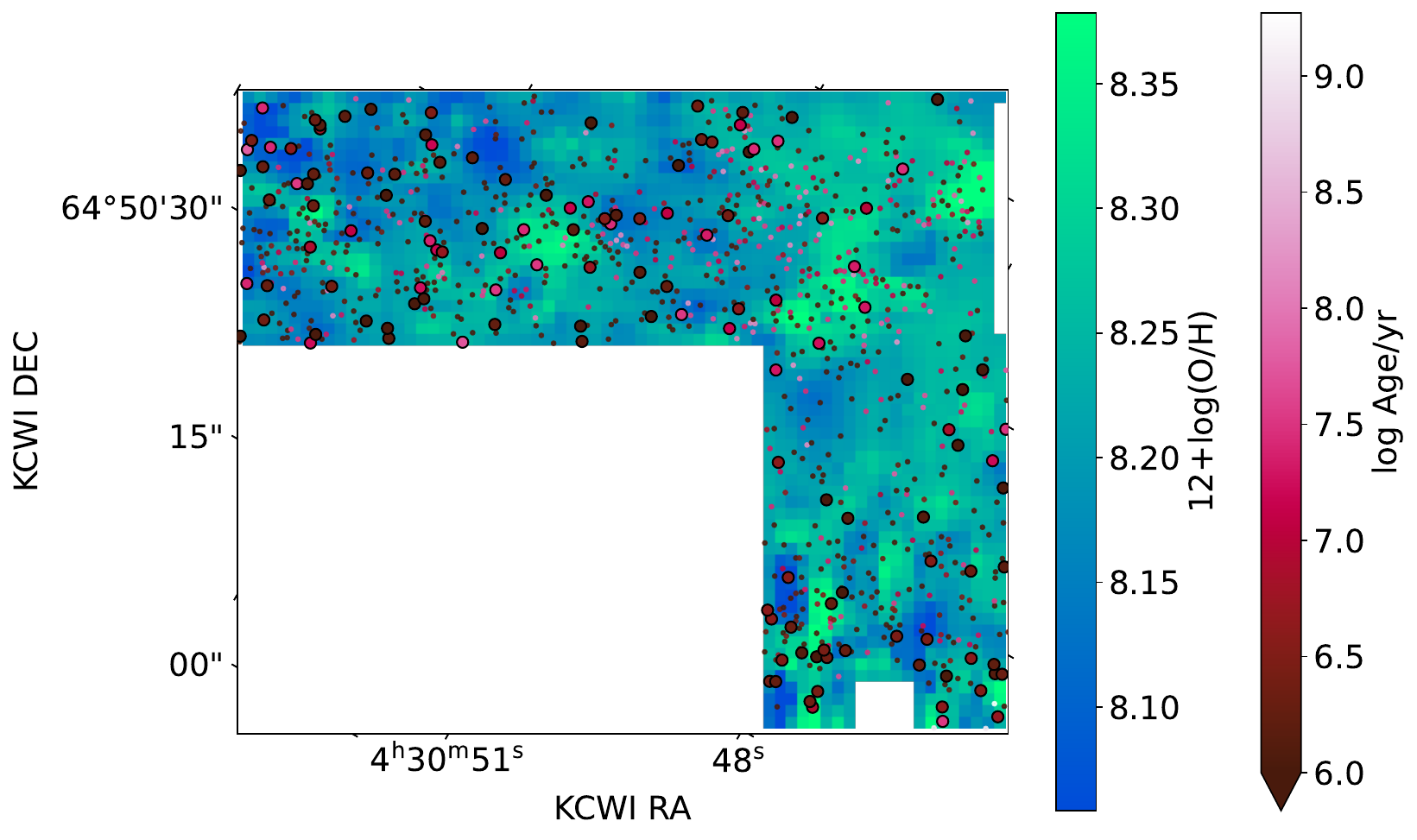}
    \includegraphics[width=0.4\linewidth]{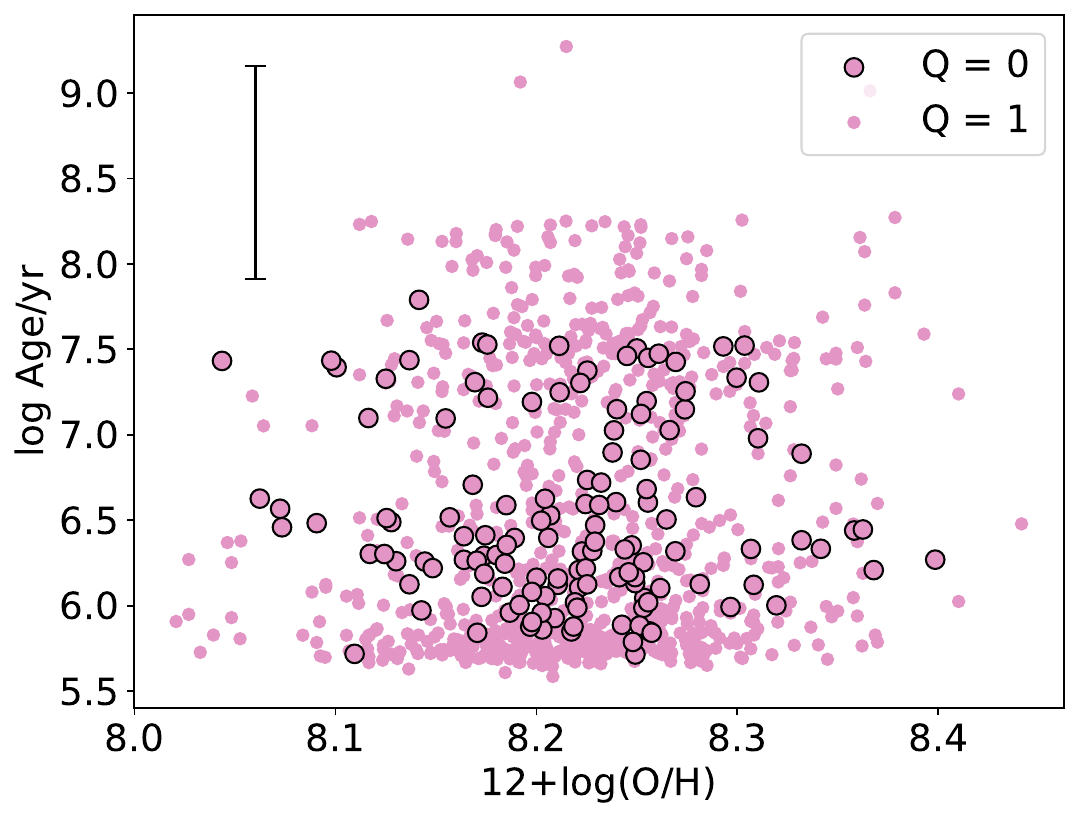}
    \caption{\textbf{Left: }As Fig.~\ref{fig:mass-metals} but showing the inferred cluster age in place of the mass.}
    \label{fig:age-metals}
\end{figure*}

\begin{figure*}
    \centering
    \includegraphics[width=0.5\linewidth]{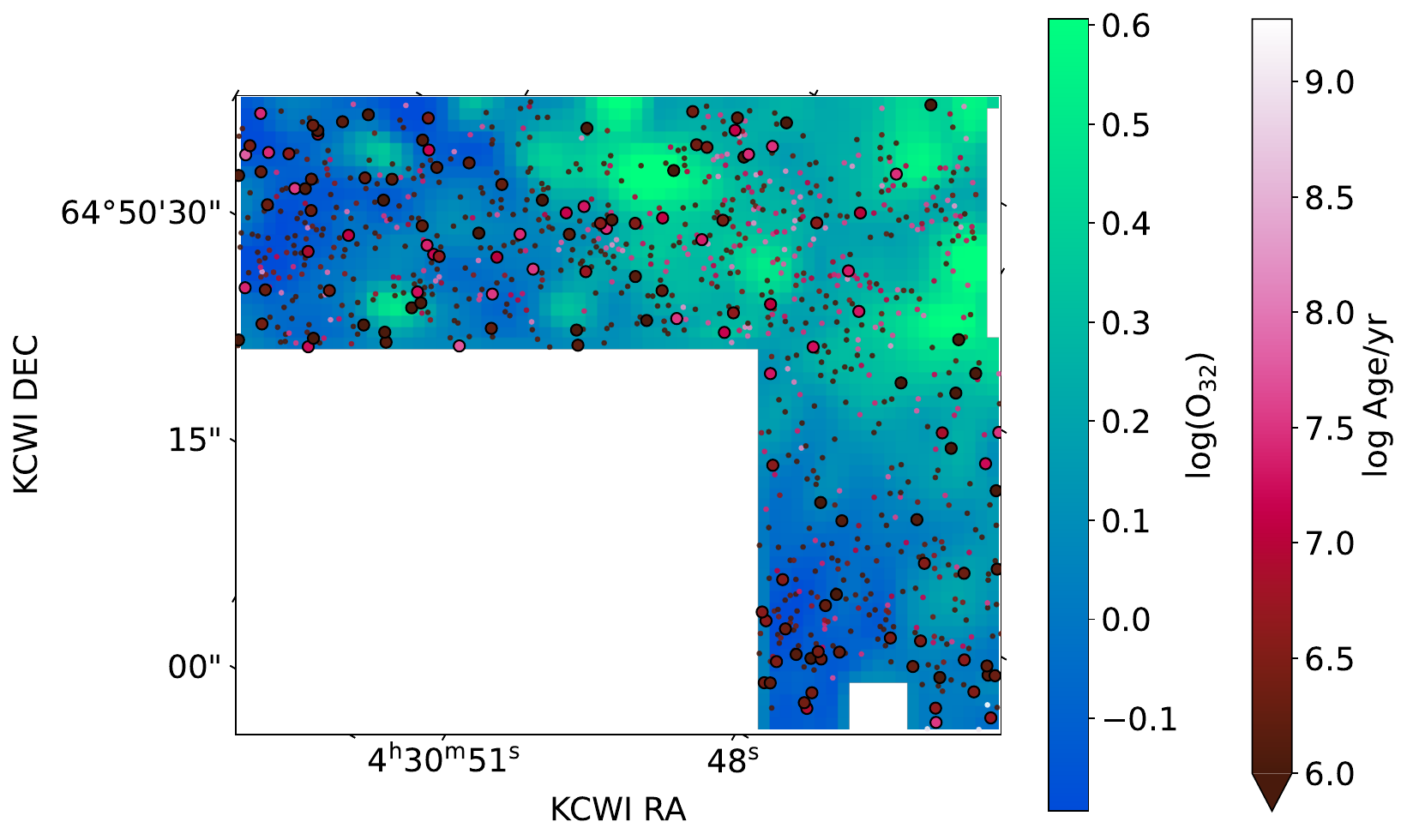}
    \includegraphics[width=0.4\linewidth]{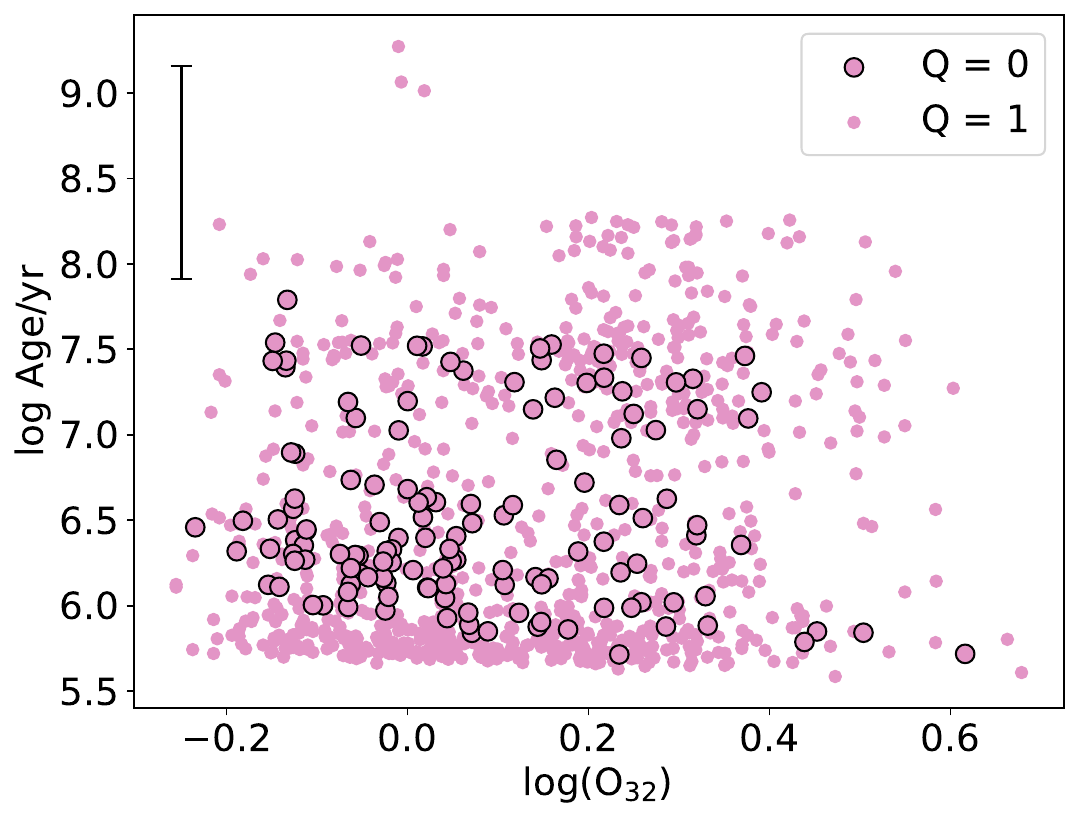}
    \caption{\textbf{Left: }As Fig.~\ref{fig:age-metals} but showing log(O32) in place of the 12+log(O/H) metallicity.}
    \label{fig:age-o32}
\end{figure*}

Lastly, we plot the same ISM properties against cluster age. We show the age-metallicity comparison in Fig.~\ref{fig:age-metals}. We see no apparent trend between the two paramters. Similarly for the ionisation state (Fig.~\ref{fig:age-o32}) we find no clear correlation.

\section{Discussion}\label{sec:Discussion}
\subsection{Cluster masses}
In Section \ref{sec:bayes-results} we showed tentative results indicating a star formation episode-dependent mean cluster mass, where clusters formed during earlier episodes appear to have higher masses, on average. While incompleteness at low masses—especially for older, faint clusters in crowded regions—likely contributes to some of these trends, the observed shifts in the location of the mass distribution peaks suggest that the cluster mass function may genuinely vary between star formation events. Similar environmental and temporal variations in cluster populations have been noted in other systems: for example, \citet{2009A&A...494..539L} and \citet{zhang_formation_1999} discuss how cluster mass functions and formation efficiencies can differ between quiescent and starburst environments. Previous studies of NGC~1569’s star formation history based on resolved stars \citep{angeretti_complex_2005} and cluster SED fits \citep{anders_star_2004} already established multiple recent bursts, and our results extend this picture by indicating that the mass scale of clusters formed may also change over time. If confirmed with more complete samples and detailed modelling of selection effects, this would imply that physical conditions - such as gas density, pressure, or feedback from earlier generations of clusters - modulate not only the timing but also the nature of cluster formation in NGC~1569.

\subsection{Radial trends}
The analysis reveals a clear negative correlation between cluster mass and galactocentric distance in NGC~1569. As shown in Fig.~\ref{fig:mass-dist-maps-both} and quantified in Table \ref{tab:slope-tab}, the cluster mass decreases systematically with increasing distance from SSC~B, which we adopt as the galaxy center. The trend is robust across both quality flags, and persists even when using only HST photometry, demonstrating that this is not an artifact of background contamination or blending in the KCWI data, though these effects may contribute to the systematic offset between quality flags.

Interestingly, we find a directional dependence of the radial mass gradient. That is, analysing the gradient separately along and perpendicular to the galactic disk (with the perpendicular direction being that along the star burst-driven outflow of the galaxy), we find a steeper slope perpendicular to the disk than along it. This anisotropic gradient suggests that the truncation mass of the CMF is not simply a function of galactocentric radius, but rather depends on the local gas density or pressure environment.

This finding is consistent with previous observations of radial mass gradients in other galaxies. \cite{della_bruna_stellar_2022} find that the mass of young star clusters in M83 decreases with galactocentric distance, with \citeauthor{adamo_probing_2015} reporting a correlation between the truncation mass and the local gas surface density. Our results extend this picture by demonstrating that the gradient is not purely radial, but also depends on the direction relative to the disk plane, with the steepest decline occurring along the minor axis where gas densities are lower in the channel of the outflow.

The physical interpretation of this mass gradient can be understood within the framework of feedback-regulated cluster formation. \cite{reina-campos_unified_2017} developed analytical models showing that the truncation mass of the CMF depends on the balance between the feedback timescale and the free-fall timescale of the parent molecular cloud. In lower density environments, stellar feedback acts on shorter timescales than the cloud collapse time, causing the cloud to be disrupted before it can convert a significant fraction of its mass into stars. This limits the maximum cluster mass that can form. In higher density regions, the stronger gravitational binding energy of the collapsing cloud allows it to resist disruption longer, enabling more mass to be converted into stars before feedback disperses the remaining gas. The steeper gradient perpendicular to the disk, where gas densities are lower due to the outflow and lower vertical gravitational potential, supports this interpretation: clusters forming in the lower-density gas above and below the disk are more readily disrupted, leading to lower maximum cluster masses.

The fact that we observe a lack of high-mass clusters in the outskirts even if lower-mass clusters might be undetected due to crowding effects further supports the interpretation that this is a physical truncation rather than a detection limit. If the gradient were purely due to incompleteness, we would expect to see some high-mass clusters in the outskirts, but these are notably absent.

In contrast to the clear mass gradient, we find no significant spatial trends in cluster age (Fig.~\ref{fig:age-dist-maps}) with galactocentric distance. The lack of an age gradient suggests that star formation has occurred across the galaxy over similar timescales, consistent with the multiple star formation episodes identified in Section \ref{sec:spatial_analysis} and previous work.

These results demonstrate that cluster formation in NGC~1569 is regulated by local environmental conditions. The anisotropic mass gradient, with its steeper decline along the outflow of the galaxy, provides observational evidence that feedback-mediated gas density plays a crucial role in setting the maximum cluster mass that can form in dwarf starburst galaxies.

\subsection{Connection to ISM properties}
The co-spatial IFU spectroscopy from KCWI enables us to directly investigate the relationship between stellar cluster properties and the conditions of their surrounding interstellar medium. Using the gas-phase metallicity and ionization parameter maps derived by \cite{MagdaPaper} from the same KCWI observations, we can examine how cluster formation responds to local ISM conditions.

\subsubsection{Cluster mass and gas-phase metallicity}
We find a weak positive correlation between cluster mass and the local gas-phase metallicity, as shown in Fig.~\ref{fig:mass-metals}. The highest mass clusters are preferentially located in regions of elevated metallicity, with 12+log(O/H) values reaching up to $\sim$8.35 near SSC~B, compared to the galaxy-wide average of $\sim$8.22 \citep{MagdaPaper}. While this correlation may suggest that massive cluster formation is favored in gas that has been locally pre-enriched by previous generations of star formation and feedback, metallicity itself is likely not the direct causal agent for higher cluster mass. Instead, the elevated metallicity in the vicinity of SSC~B likely results from the enrichment of the local ISM by stellar winds and supernova ejecta from SSC~B itself as well as and the surrounding massive clusters over multiple star formation episodes. 

\subsubsection{Cluster mass and degree of ionization}
The relation between cluster mass and the ionisation parameter of the surrounding gas, proxied by log(O32) = log([\oIII]$\lambda$5007/[\oII]$\lambda$3727), reveals a positive correlation (Fig.~\ref{fig:mass-O32}). Higher mass clusters (and the hot stars within them) are associated with higher ionization parameters, indicating that more massive clusters produce stronger radiation fields capable of ionizing their surroundings to higher degrees. The relationship between cluster mass and ionizing photon flux $Q_{0}$ is evident in the synthetic cluster library (Fig.~\ref{fig:mag-prop}), where we see a positive correlation between mass and $Q_{0}$. The observed correlation with log(O32) thus reflects the direct connection between cluster mass and the production of ionizing radiation.

However, we note that the ionization parameter also depends on the gas density and geometry of the ionised gas, that is, on the optical depth of the regions. The correlation we observe may therefore be influenced by the fact that massive clusters tend to form in denser gas regions, where the higher gas density could also contribute to higher ionization parameters for a given ionizing photon flux. Nevertheless, the spatial correlation between high cluster masses and high log(O32) values supports the interpretation that massive clusters are the primary drivers of the high ionization state in these regions.

\subsubsection{Cluster age and ISM properties}
In contrast to the correlations observed for cluster mass, we find no significant trends between cluster age and either metallicity or ionisation parameter (Fig.~\ref{fig:age-metals}). The lack of an age-metallicity correlation suggests that clusters of all ages are forming in gas with similar metallicity, consistent with the idea that the metallicity gradient is primarily spatial rather than temporal. This could indicate that enrichment occurs on timescales longer than the age spread of our cluster sample, or that enriched gas is efficiently mixed throughout the star-forming regions.

The absence of a correlation between age and ionization parameter is somewhat surprising, as younger clusters would be expected to contain more massive stars and produce stronger radiation fields. However, this may reflect the fact that log(O32) traces the current ionization state, which could be influenced by the cumulative effect of multiple clusters in crowded regions, or by the time delay between cluster formation and the establishment of a photoionization equilibrium in the surrounding gas. Additionally, the age resolution of our analysis, particularly for clusters younger than $\sim$3 Myr, may limit our ability to detect age-dependent trends in ionization.

\subsubsection{Implications for cluster formation}

Taken together, these correlations provide a picture of cluster formation in NGC~1569 regulated by local ISM conditions. Massive clusters form preferentially in regions of high gas density (where super star clusters like SSC~B can form, and as evidenced by the radial mass gradient), high metallicity, and produce strong radiation fields that maintain high ionisation parameters in their surroundings. This suggests a positive feedback loop: massive clusters form in dense, enriched gas, enrich their surroundings further through stellar winds and supernovae, and create conditions that may favor subsequent massive cluster formation.

The spatial concentration of these favorable conditions near SSC~B, where we observe the highest cluster masses, highest metallicities, and highest ionisation parameters, indicates that this region represents a particularly active site of cluster formation and feedback. The fact that this region has sustained multiple generations of massive cluster formation (as evidenced by the range of cluster ages) suggests that feedback from massive clusters can create and maintain conditions conducive to further massive cluster formation, at least in the dense central regions of dwarf starburst galaxies.

These results demonstrate the power of combining high-resolution photometry with IFU spectroscopy to directly link cluster demographics to the physical conditions of their birth environments. The correlations we observe provide observational support for theoretical models in which cluster formation is regulated by a complex interplay between gas density, metallicity, and feedback processes.

\section{Summary}\label{sec:Summary}
Using a combination of high-resolution HST photometry and IFU spectro-photometric data from KCWI covering the wavelength range from $\sim$3600~\AA\ to $\sim$9600~\AA, we have derived star cluster physical properties in the nearby dwarf starburst galaxy NGC~1569 using Bayesian inference and synthetic photometry of stochastically sampled star clusters.
Our key findings are as follows:
\begin{itemize}
    \item In the wavelength range covered by the HST and KCWI data, the optimal combination of filters in terms of recovering reliable cluster parameters vs. number of clusters that could not be fit at all is HST/ACS combined with the three KCWI broadband filters.
    \item Further, we show that additional bluer coverage with HST and NIR coverage with JWST/NIRCam - that is, with coverage from the UV to $\sim$5 $\mu$m is crucial to break the degeneracy between age and colour.
    \item We successfully recover two known episodes of star formation reported in the literature, and, while caveats need to be taken into consideration, we also report a younger episode of star formation younger than 10$^{6.5}$ yr.
    \item The distribution of cluster masses differs between young, intermediate, and old populations, with statistically significant differences between the youngest and older age bins. Within caveats, this tentatively suggests that NGC~1569’s star formation history is not only episodic but may also involve variations in the typical masses of clusters formed during different bursts, consistent with environmental modulation of the cluster mass function.
    \item We find a negative trend of cluster mass with radial distance from the super star cluster SSC~B, as well as a steeper negative trend along the galactic outflow of NGC~1569. We interpret this as feedback-mediated gas density enabling high-mass cluster formation close to the main source of feedback, and truncating cluster masses in the lower density gas, in particular along the outflow cavity.
    \item Together with the previous, we find a weak positive trend of cluster mass with gas-phase metallicity, which possibly suggests that massive clusters form in gas that has been locally pre-enriched by sustained star formation and feedback. This is further supported by the fact that the highest metallicities are observed close to SSC~B where we also observe higher cluster masses.
    \item We also find a positive trend of cluster mass with the degree of ionisation of the surrounding gas, likely reflecting the fact that higher mass clusters contain higher mass stars capable of producing stronger radiation fields.
\end{itemize}

The methods and analyses used in this paper demonstrate that optical IFU spectroscopy and co-spatial high-resolution photometry, combined with stochastic cluster population modelling, enable us to link cluster demographics directly to feedback-regulated star formation, providing a powerful avenue for dissecting the physics of dwarf starbursts. Taken together, these results contribute to providing a picture in which star cluster formation in dwarf starburst galaxies is regulated by a spatially and directionally varying interplay between gas density, feedback effectiveness, and stochasticity, and show that resolved cluster populations can be used as quantitative tracers of how feedback sculpts the ISM and limits cluster growth in low-mass galaxies across the local Universe.

\section*{Acknowledgements}
This work was supported by Science and Technology Facilities Council grant ST/Y509346/1.
Some of the data presented herein were obtained at Keck Observatory, which is a private 501(c)3 non-profit organization operated as a scientific partnership among the California Institute of Technology, the University of California, and the National Aeronautics and Space Administration. The Observatory was made possible by the generous financial support of the W. M. Keck Foundation. The authors wish to recognize and acknowledge the very significant cultural role and reverence that the summit of Maunakea has always had within the Native Hawaiian community. We are most fortunate to have the opportunity to conduct observations from this mountain.\\
This work made use of Astropy:\footnote{http://www.astropy.org} a community-developed core Python package and an ecosystem of tools and resources for astronomy \citep{astropy:2013, astropy:2018, astropy:2022}\\

%%%%%%%%%%%%%%%%%%%%%%%%%%%%%%%%%%%%%%%%%%%%%%%%%%
\section*{Data Availability}

All raw KCWI data files are accessible in the Keck Observatory Archive\footnote{https://www2.keck.hawaii.edu/koa/public/koa.php}.\\
The HST data were obtained using ESASky, developed by the ESAC Science Data Centre (ESDC) team and maintained alongside other ESA science mission's archives at ESA's European Space Astronomy Centre (ESAC, Madrid, Spain).

%%%%%%%%%%%%%%%%%%%% REFERENCES %%%%%%%%%%%%%%%%%%

% The best way to enter references is to use BibTeX:

\bibliographystyle{mnras}
\bibliography{bib} % if your bibtex file is called example.bib

% Alternatively you could enter them by hand, like this:
% This method is tedious and prone to error if you have lots of references
%\begin{thebibliography}{99}
%\bibitem[\protect\citeauthoryear{Author}{2012}]{Author2012}
%Author A.~N., 2013, Journal of Improbable Astronomy, 1, 1
%\bibitem[\protect\citeauthoryear{Others}{2013}]{Others2013}
%Others S., 2012, Journal of Interesting Stuff, 17, 198
%\end{thebibliography}

%%%%%%%%%%%%%%%%%%%%%%%%%%%%%%%%%%%%%%%%%%%%%%%%%%

%%%%%%%%%%%%%%%%% APPENDICES %%%%%%%%%%%%%%%%%%%%%

\appendix

\section{KCWI errors}\label{app:err_comp}
\begin{figure}
    \centering
    \includegraphics[width=0.8\linewidth]{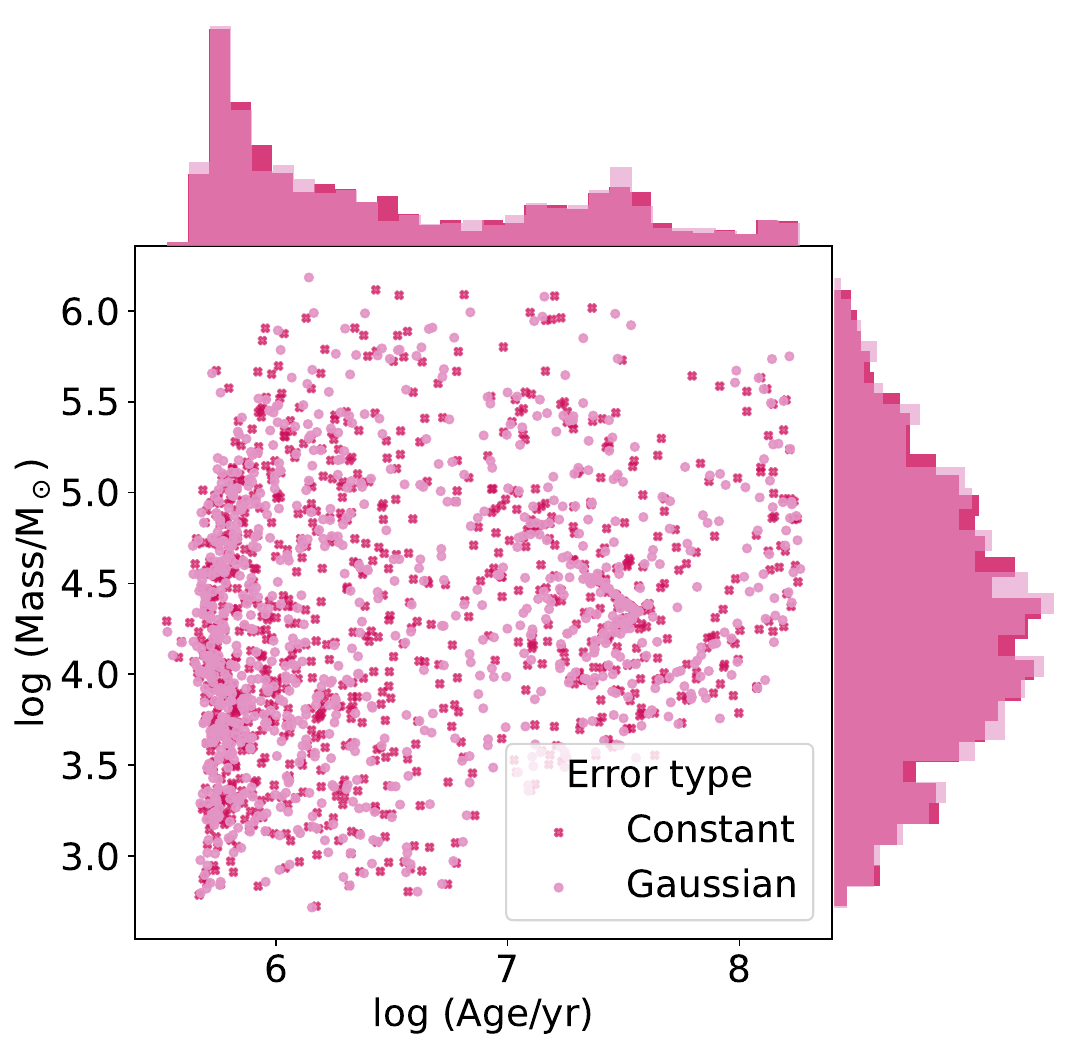}
    \caption{A plot comparing the results of the fitting using two different types of errors. The x-axis shows the logarithmic age of the clusters, and the y-axis the logarithm of the mass in Solar masses. The histograms in the margins show the distribution of the same properties. The magenta crosses show the results when a constant 20\% error is used, while the pink points show results from using errors sampled from a Gaussian with mean 20\% and  $\sigma=5\%$. Although there are slight differences between individual values, the overall distributions are consistent between error types.}
    \label{fig:errortype_comparison}
\end{figure}

\begin{figure}
    \centering
    \includegraphics[width=0.8\linewidth]{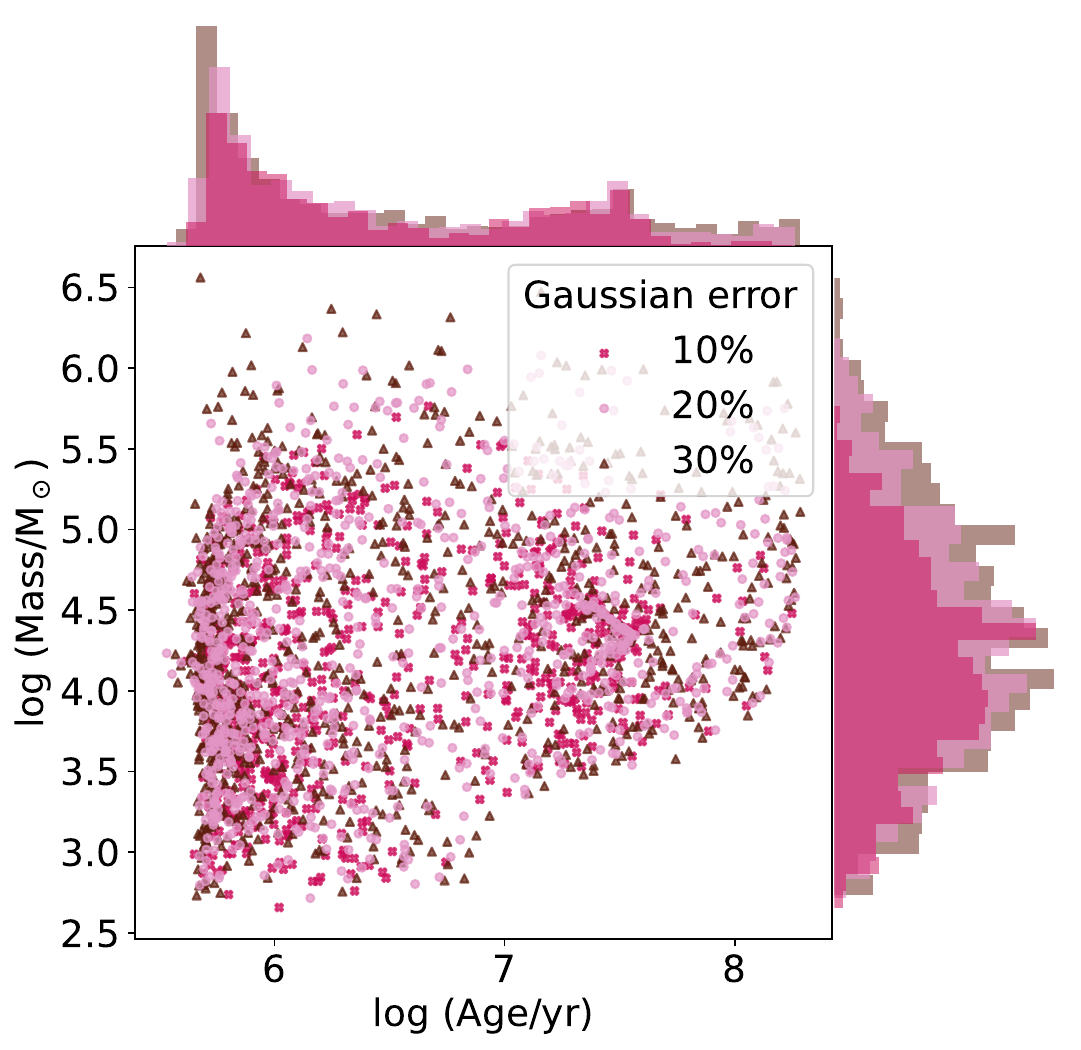}
    \caption{Same as figure \ref{fig:errortype_comparison} except comparing the size of the error rather than the type of error. The pink points show the results when errors are sampled from a Gaussian with mean 20\%, the magenta crosses have errors sampled from a Gaussian with mean of 10\%, the brown triangles show results using Gaussian with mean 30\%. All Gaussians have a $\sigma=5\%$. Despite differences between individual values, the overall distributions are largely consistent between error sizes.}
    \label{fig:errorsize_comparison}
\end{figure}

To show that the exact errors are not necessary for this work, we compare results when using different values for the error of the KCWI data. A visual comparison is shown in figure \ref{fig:errortype_comparison}, where a constant 20\% error is compared to errors sampled from a Gaussian with a mean of 20\% and $\sigma=5\%$. and figure \ref{fig:errorsize_comparison}, where a Gaussian with a mean of 20\% is compared with one with a mean of 10\% and another with a mean of 30\%. The values shown here are simply the median of the marginal 1D distributions of mass and age for each cluster.

Comparing the Gaussian sampled errors with the constant error, we see that using the constant errors rather than the Gaussians, has very little effect on the fitted results, although the results don't match exactly, the overall distribution of points remains the same. The variations in the obtained properties are smaller than the spread of the posterior PDFs.

Comparing the  Gaussians, we see that although there are small differences in the returned values, the overall distributions remain largely the same and there seems to be no reason to pick one over the other. Increasing the errors from 10\% to 20\%, going from 28\% to 11\% having no constraints on their propertis, but increasing from 20\% to 30\% gives us an additional 4\% of clusters being constrained (7\% not fit), but does also increase the 

%%%%%%%%%%%%%%%%%%%%%%%%%%%%%%%%%%%%%%%%%%%%%%%%%%

% Don't change these lines
\bsp	% typesetting comment
\label{lastpage}
\end{document}